\documentclass[12pt,preprint]{aastex}
 
\begin{document}

\title{A $^{12}$CO, $^{13}$CO, and C$^{18}$O Survey of Infrared Dark Clouds}

\author{Fujun Du, Ji Yang}
\affil{Purple Mountain Observatory, Chinese Academy of Sciences}
\affil{2 West Beijing Road, Nanjing 210008}
\email{fjdu@pmo.ac.cn}
\email{jiyang@pmo.ac.cn}

\begin{abstract}
InfraRed Dark Clouds (IRDCs) are extinction features against the Galactic infrared background, mainly in mid-infrared band.
Recently they were proposed to be potential sites of massive star formation.
In this work we have made $^{12}$CO, $^{13}$CO, and C$^{18}$O (J=1$\rightarrow$0)\ survey of 61 IRDCs, 
52 of which are in the first Galactic quadrant, selected from a catalog given by Simon et al. (2006), 
while the others are in the outer Galaxy, selected by visually inspecting the MSX images. 
Detection rates in the three CO lines are 90\%, 71\%, and 62\% respectively. 
The distribution IRDCs in the first Galactic quadrant is consistent with the 5 kpc molecular ring picture,
while slight trace of spiral pattern can also be noticed, which needs to be further examined.
The IRDCs have typical excitation temperature of 10 K and typical column density of several $10^{22}$ cm$^{-2}$.
Their typical physical size is estimated to be several $pc$s using angular sizes from the Simon catalog.
Typical volume density and typical LTE mass are $\sim$5000 cm$^{-3}$ and $\sim$5000 M$_\odot$ respectively.
The IRDCs are in or near virial equilibrium.
The properties of IRDCs are similar to those of star forming molecular clumps,
and they seem to be intermediate between giant molecular clouds and Bok globules,
thus they may represent early stages of massive star formation.
\end{abstract}


\section{Introduction}
The infrared dark clouds (IRDCs), revealed by the Midcourse Space
Experiment (MSX), are dark extinction features against the Galactic
mid-infrared background \citep{Egan1998a},
usually with filamentary or compact shapes, mostly at 8.6 -- 10.6 ~$\mu$m.
High opacity suggests the IRDCs might have high gas column density.
\citet{Carey1998a, Carey2000a} showed that the IRDCs have column densities as high as $10^{23}$ cm$^{-2}$,
gas density $\sim$$10^5$ -- $10^6$ cm$^{-3}$,
with distances to the sun $\sim$1 -- 8 kpc, and diameters $\sim$0.4 -- 15 pc.
The masses of the clouds they observed are estimated to be from tens of to thousands of solar masses,
while the temperatures are typically 10 -- 25 K.
Based on millimeter-to-mid-IR continuum observation of an MSXDC,
\citet{Rathborne2005a} found very high luminosities, 9000 -- 32,000 L$_\odot$, of the cores,
and they concluded that it must be forming massive stars.
\citet{Ragan2006a} mapped 41 IRDCs in N$_2$H$^+$ 1$\rightarrow$0, CS 2$\rightarrow$1, and C$^{18}$O 1$\rightarrow$0.
They found that different species often show striking differences in morphologies,
which were attributed to differences in evolutionary state and/or the
presence of undetected, deeply embedded protostars.
Average mass of the clouds is estimated to be $\sim$2500 M$_\odot$ using N$_2$H$^+$ observations,
which is consistent with the previous studies of massive star-forming regions.
The typical line width of the clouds they observed is 2 -- 3 km s$^{-1}$.
\citet{Rathborne2006a} found that the mass spectrum of IRDC cores
derived from millimeter maps is consistent with the stellar IMF.
Assuming each core will form a single star, they concluded that the
majority of these cores will form OB stars.
The IRDC cores are similar to the hot cores associated with individual,
young high-mass stars, except that they are much colder, thus \citet{Rathborne2006a}
suggested that the IRDCs represent an earlier evolutionary phase in
massive star formation, and they may be cold precursors to star clusters.
A water maser survey of 140 IRDC compact cores of \citet{Wang2006a}
revealed that the detection rate of H$_2$O masers for higher mass
cores is significantly higher than that of lower mass cores.
They suggest that the most massive IRDC cores without H$_2$O maser may be
at an earlier stage than the protostellar phases.
A survey in\ $^{13}$CO by \citet{Simon2006b} established kinematic
distances to 313 IRDCs. They derived typical sizes of $\sim$5 pc, peak
column densities of $\sim$10$^{22}$ cm$^{-2}$, LTE masses of $\sim$5$\times10^3$
M$_\odot$, and volume-averaged H$_2$ densities of $\sim$2$\times10^3$ cm$^{-3}$.
\citet{Beuther2007a} discovered a protostar in an IRDC, which will probably become a massive star.
These previous studies suggest that IRDCs can be ideal candidates to study the
initial conditions and early stages of massive star formation.

An IRDC catalog has been published by \citet{Simon2006a}.
Clouds in this catalog are in the first and fourth quadrant of the
Galaxy. The authors used the MSX A band data to construct a background model,
and then searched for regions with rapid decrease with respect to this background.
The IRDC candidates are defined by contiguous regions
bounded by closed contours of 2$\sigma$ decremental contrast threshold.
The catalog contains 10,931 IRDCs, and 12,774 cores, with reliability estimated to be 82\%.
The distribution of IRDCs is in good correspondence with the diffuse Galactic infrared background.
There are more IRDCs toward the star-forming regions, the spiral arm tangents, and the 5 kpc molecular ring.

As part of our campaign to search for candidates of massive star forming
clouds, we selected a sample of 61 clouds mainly from the IRDC
catalog given by \citet{Simon2006a} and observed them in $J=1\rightarrow0$
lines of $\rm ^{12}CO$, $\rm ^{13}CO$, and $\rm C^{18}O$ in single
point mode. In this article we report the results of this survey.

\section{Source Selection}
Our sources are mainly selected from the Simon catalog of IRDCs \citep{Simon2006a}.
We first found out those IRDCs with contrast greater than 0.4,
area larger than $1'\times1'$, and declination greater than about $-22^\circ$. 
Here ``contrast'' is defined as $\rm (Background-Image)/Background$ \citep{Simon2006a}.
After this first round selection, there are 209 sources left.
Then we divided the sources into groups by a 2 degrees step in Galactic longitude,
and selected from each group one or more ``representatives''
with the most prominent extinction features and relatively well-defined boundaries by visually inspecting the MSX images.
After this step, there are 52 sources left, which comprise the major part of our survey.
Except for 9 sources, whose Galactic coordinates are taken to be the first or second peak positions,
the coordinates of the other objects are taken to be the cloud centroid position from the Simon catalog.

Besides these, we also selected 9 supplementary IRDC candidates against the bright mid-infrared background
in the second and third Galatic quadrant by visual inspection of the MSX images.
Their coordinates are taken to be the estimated extinction peaks.
The identification of IRDC in the outer Galaxy is difficult,
as the background emission is weaker and the MSX image quality is lower in these regions than in the inner Galaxy.
We hope to get some hint on the existence and/or distribution of IRDCs in the outer Galaxy from the enrollment of these 9 targets.

\begin{figure}[htbp]
\plotone{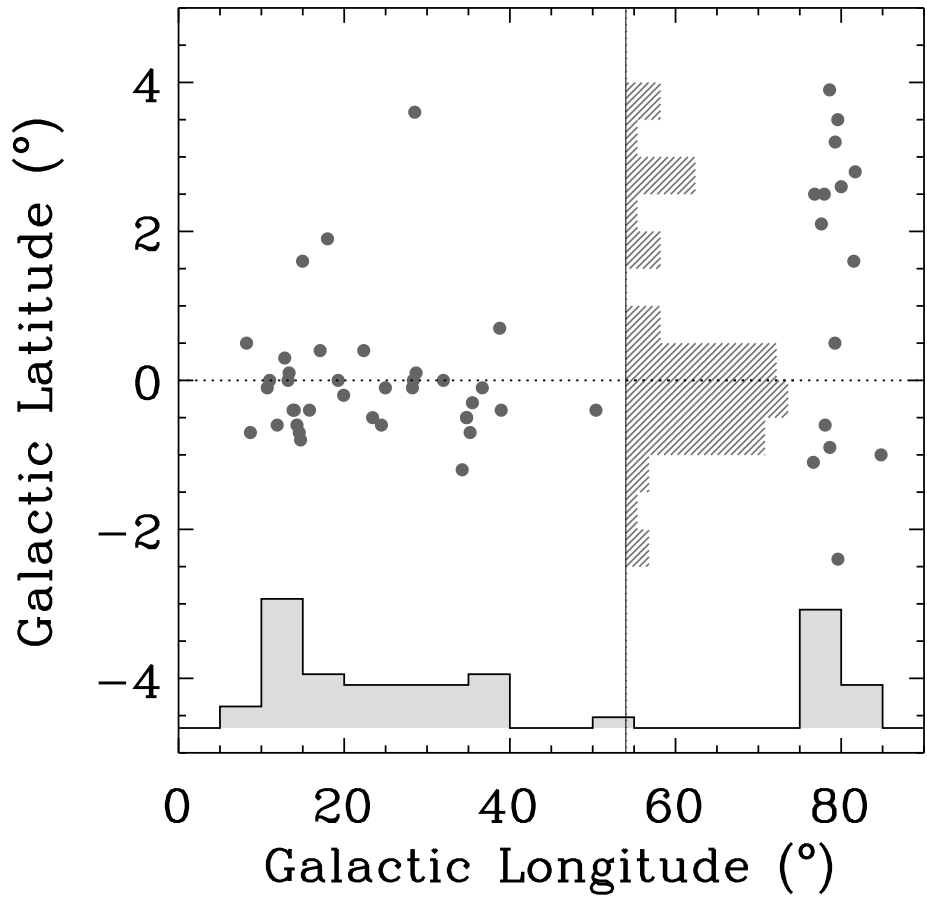}
\figcaption{Distribution of our sample in Galactic coordinates, with histograms of Galactic longitudes and latitudes overplotted.\label{fig:Gala_distri}}
\end{figure}

\begin{figure}[htbp]
\plottwo{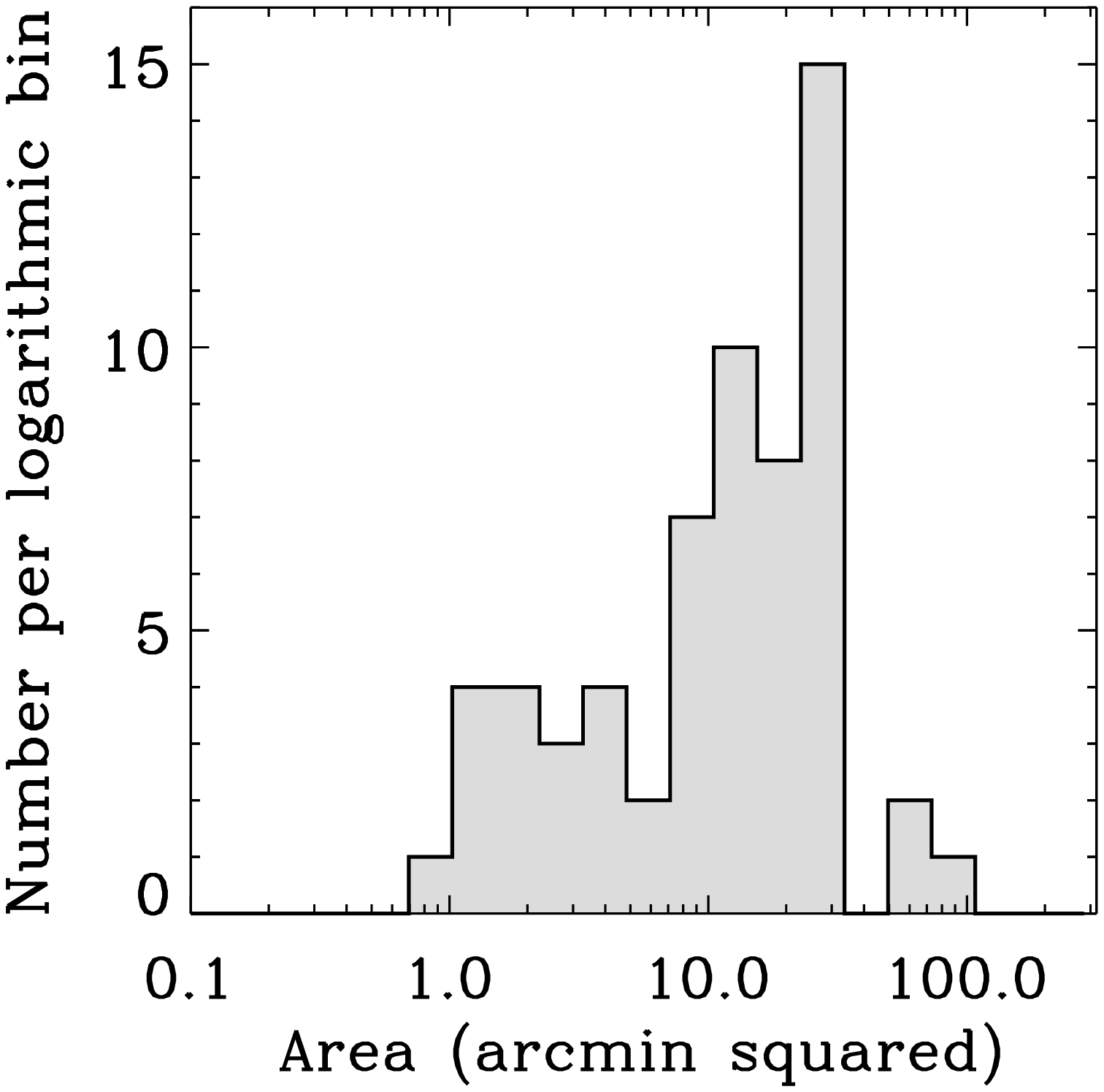}{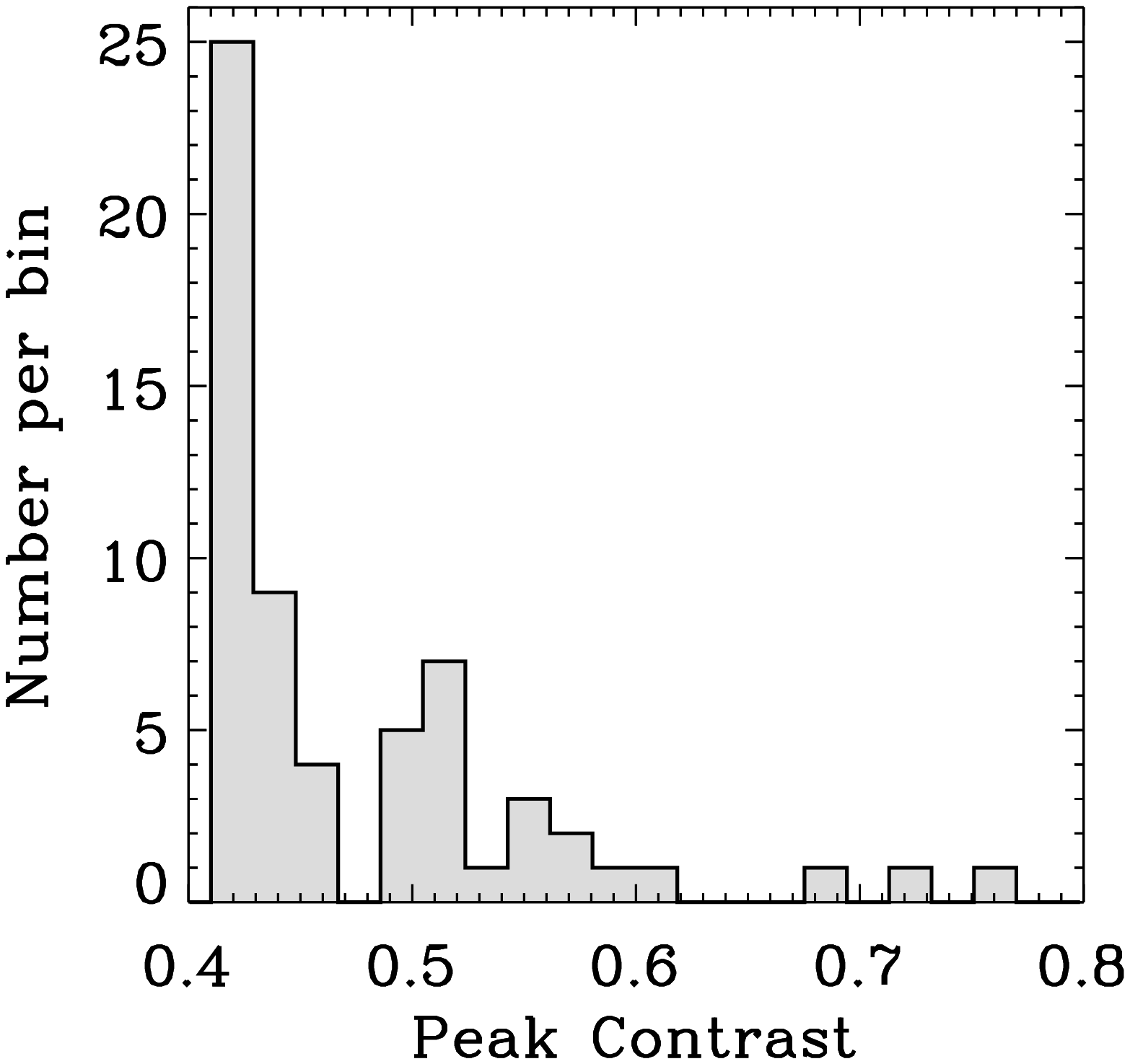}
\figcaption{Histogram of IRDC angular area and peak contrast of our sample.
Notice that the abscissa of the histogram of area is in log scale. \label{fig:area_pc_distri}}
\end{figure}

In Table \ref{tab:sname_coordinates} we list the source names, Galactic coordinates, 
and Equatorial coordinates used during the observation.
Figure \ref{fig:Gala_distri} shows the Galactic distribution of our
sample; the sources out of the first quadrant are not plotted in this
figure. We notice that almost none of our sources fall in the range
between $l=50^\circ$ and $l=70^\circ$, which clearly inherits a feature of
the Simon catalog (cf. Figure 5 of \citet{Simon2006a}).
Figure \ref{fig:area_pc_distri} shows histograms of the area and peak contrast of our sample,
with data taken from \citet{Simon2006a}.

\clearpage
\pagestyle{plaintop}
\setlength{\voffset}{0mm}
\begin{deluxetable}{r c c c c c}\tablecolumns{6} \tablewidth{0pc} 
\tabletypesize{\normalsize}
\tablecaption{List of source names, Galactic coordinates and Equatorial coordinates.
\label{tab:sname_coordinates}}
\tablehead{
\colhead{} & \colhead{} & \multicolumn{4}{c}{C{\sc oordinates}} \\
\cline{3-6}\\
\colhead{} & \colhead{N{\sc ame}} & \colhead{$l$} & \colhead{$b$} & \colhead{R. A.} & \colhead{D{\sc ecl}.}\\
\colhead{} & \colhead{MSXDC} & \colhead{(deg)} & \colhead{(deg)} & \colhead{(J2000)} & \colhead{(J2000)} \\
\colhead{(1)} & \colhead{(2)} & \colhead{(3)} & \colhead{(4)} & \colhead{(5)} & \colhead{(6)}
}
\startdata 

 1 & G008.21+00.47\tablenotemark{a}  &  008.211  &  +0.514  &  18:02:05  &  $-$21:35:52\\
 2 & G008.67$-$00.70  &  008.677  &  $-$0.709  &  18:07:40  &  $-$21:47:27\\
 3 & G010.71$-$00.16  &  010.710  &  $-$0.167  &  18:09:51  &  $-$19:45:00\\
 4 & G010.99$-$00.07  &  010.994  &  $-$0.074  &  18:10:05  &  $-$19:27:23\\
 5 & G011.87$-$00.62\tablenotemark{a}  &  011.909  &  $-$0.622  &  18:13:59  &  $-$18:55:03\\
 6 & G012.81+00.36  &  012.815  &  +0.368  &  18:12:09  &  $-$17:38:52\\
 7 & G013.22$-$00.07  &  013.224  &  $-$0.079  &  18:14:37  &  $-$17:30:11\\
 8 & G013.34+00.17  &  013.344  &  +0.180  &  18:13:55  &  $-$17:16:25\\
 9 & G013.82$-$00.48  &  013.825  &  $-$0.483  &  18:17:19  &  $-$17:10:01\\
10 & G013.97$-$00.43  &  013.978  &  $-$0.437  &  18:17:27  &  $-$17:00:37\\
11 & G014.29$-$00.66  &  014.297  &  $-$0.666  &  18:18:55  &  $-$16:50:16\\
12 & G014.56$-$00.78  &  014.569  &  $-$0.785  &  18:19:54  &  $-$16:39:16\\
13 & G014.72$-$00.88\tablenotemark{b}  &  014.731  &  $-$0.897  &  18:20:38  &  $-$16:33:52\\
14 & G014.97+01.60  &  014.974  &  +1.606  &  18:11:57  &  $-$15:09:35\\
15 & G015.80$-$00.40  &  015.809  &  $-$0.409  &  18:20:57  &  $-$15:23:01\\
16 & G017.09+00.45  &  017.094  &  +0.460  &  18:20:17  &  $-$13:50:27\\
17 & G017.98+01.97  &  017.982  &  +1.975  &  18:16:32  &  $-$12:20:30\\
18 & G019.27+00.07  &  019.271  &  +0.074  &  18:25:52  &  $-$12:05:57\\
19 & G019.92$-$00.29  &  019.927  &  $-$0.293  &  18:28:27  &  $-$11:41:22\\
20 & G022.35+00.41  &  022.357  &  +0.416  &  18:30:29  &  $-$09:12:28\\
21 & G023.42$-$00.52  &  023.430  &  $-$0.525  &  18:35:52  &  $-$08:41:25\\
22 & G024.49$-$00.69  &  024.491  &  $-$0.698  &  18:38:27  &  $-$07:49:38\\
23 & G025.04$-$00.20\tablenotemark{a}  &  024.974  &  $-$0.166  &  18:37:27  &  $-$07:09:14\\
24 & G028.23$-$00.19  &  028.235  &  $-$0.191  &  18:43:32  &  $-$04:16:00\\
25 & G028.37+00.07\tablenotemark{a}  &  028.341  &  +0.058  &  18:42:51  &  $-$04:03:30\\
26 & G028.51+03.60  &  028.514  &  +3.609  &  18:30:31  &  $-$02:16:35\\
27 & G028.67+00.13  &  028.677  &  +0.132  &  18:43:12  &  $-$03:43:33\\
28 & G031.97+00.07  &  031.976  &  +0.071  &  18:49:26  &  $-$00:49:06\\
29 & G034.24$-$01.25  &  034.246  &  $-$1.252  &  18:58:17  &  +00:35:54\\
30 & G034.77$-$00.55  &  034.771  &  $-$0.557  &  18:56:47  &  +01:22:57\\
31 & G034.77$-$00.55\tablenotemark{a}  &  034.782  &  $-$0.558  &  18:56:48  &  +01:23:33\\
32 & G035.19$-$00.72\tablenotemark{a}  &  035.201  &  $-$0.726  &  18:58:10  &  +01:41:17\\
33 & G035.39$-$00.33\tablenotemark{a}  &  035.479  &  $-$0.303  &  18:57:10  &  +02:07:46\\
34 & G036.67$-$00.11  &  036.673  &  $-$0.120  &  18:58:42  &  +03:16:27\\
35 & G038.77+00.78  &  038.772  &  +0.789  &  18:59:18  &  +05:33:24\\
36 & G038.95$-$00.47  &  038.952  &  $-$0.475  &  19:04:09  &  +05:08:17\\
37 & G050.39$-$00.41  &  050.395  &  $-$0.414  &  19:25:36  &  +15:17:39\\
38 & G076.64$-$01.13  &  076.642  &  $-$1.133  &  20:30:18  &  +37:17:25\\
39 & G076.79+02.59  &  076.794  &  +2.596  &  20:15:07  &  +39:30:50\\
40 & G077.61+02.10  &  077.610  &  +2.102  &  20:19:37  &  +39:56:48\\
41 & G077.95+02.59  &  077.957  &  +2.598  &  20:18:30  &  +40:30:45\\
42 & G078.06$-$00.67  &  078.064  &  $-$0.677  &  20:32:42  &  +38:42:24\\
43 & G078.60+03.92  &  078.602  &  +3.922  &  20:14:33  &  +41:47:04\\
44 & G078.62$-$00.93  &  078.629  &  $-$0.932  &  20:35:29  &  +39:00:28\\
45 & G079.24+00.52\tablenotemark{a}  &  079.244  &  +0.529  &  20:31:17  &  +40:22:18\\
46 & G079.28+03.25  &  079.280  &  +3.259  &  20:19:32  &  +41:58:36\\
47 & G079.58+03.59  &  079.584  &  +3.591  &  20:18:59  &  +42:24:52\\
48 & G079.60$-$02.49  &  079.606  &  $-$2.499  &  20:44:55  &  +38:49:28\\
49 & G080.00+02.67  &  080.002  &  +2.680  &  20:24:19  &  +42:14:26\\
50 & G081.52+01.60  &  081.522  &  +1.608  &  20:33:52  &  +42:50:43\\
51 & G081.69+02.85  &  081.694  &  +2.856  &  20:28:53  &  +43:43:14\\
52 & G084.81$-$01.09  &  084.814  &  $-$1.095  &  20:56:45  &  +43:44:13\\
53 & G093.14+02.71  &  093.144  &  +2.713  &  21:13:11  &  +52:28:37\\
54 & G110.97$-$00.85  &  110.970  &  $-$0.854  &  23:14:17  &  +59:44:30\\
55 & G111.04$-$00.64  &  111.041  &  $-$0.641  &  23:14:12  &  +59:58:00\\
56 & G133.28+00.21  &  133.283  &  +0.218  &  02:19:22  &  +61:18:38\\
57 & G173.38+02.57  &  173.384  &  +2.579  &  05:39:31  &  +35:55:25\\
58 & G189.97+00.45  &  189.972  &  +0.457  &  06:09:44  &  +20:23:58\\
59 & G190.12+00.45  &  190.128  &  +0.457  &  06:09:44  &  +20:23:58\\
60 & G206.30$-$02.01  &  206.304  &  $-$2.012  &  06:32:08  &  +04:58:14\\
61 & G206.91$-$02.45  &  206.914  &  $-$2.453  &  06:29:50  &  +05:06:44\\
\enddata
\tablenotetext{a}{Fisrt peak postion is adopted.}
\tablenotetext{b}{Second peak postion is adopted.}
\end{deluxetable}
\clearpage

\clearpage

\section{Observation}
Our observations were taken during the observation season between 2006 and 2007
with the 13.7 m radio telescope in Delingha, China, which is operated by Purple Mountain Observatory.

The SIS receiver used during the observation works in a double sideband mode.
The three CO lines are received simultaneously, with $^{12}$CO in the upper sideband and 
$^{13}$CO and C$^{18}$O in the lower sideband \citep{Zuo2004a}.
The total system temperature is typically about 200\ --\ 300 K for the CO observations. 
The backend AOS spectrometers have a resolution of 0.142 MHz (corresponding to a velocity resolution of
0.37 km s$^{-1}$) for $^{12}$CO, and 0.042 MHz (0.11 km s$^{-1}$) for $^{13}$CO\ and C$^{18}$O, with
bandwidth being 378 km s$^{-1}$, 116 km s$^{-1}$, and 118 km s$^{-1}$ respectively.
The observations were taken in position switch mode.
Data were calibrated using the standard chopper wheel method.

The absolute pointing of the telescope has an RMS
accuracy of about $10''$ during the observations.
The main beam width is about $1^\prime$.
The stability of the receiver and the pointing were checked every two hours by observing stand sources.
The observed intensities were corrected for the main beam efficiency,
which was obtained from observations of standard sources and planets.
Taking into account of the uncertainty in the main beam efficiency, 
and the fact that the coupling efficiency is different for each object,
the observed intensities of our sources can be treated as certain within a factor of 1.5.
Data are processed using the CLASS package of GILDAS software.

\section{Results}
A sample of 61 IRDCs, most of whose coordinates are taken to be the cloud coordinates in the Simon catalog,
have been surveyed in $^{12}$CO, $^{13}$CO, and C$^{18}$O.
Due to different available observation time and varying weather
conditions for each target, the resulting RMS noises for each source as well as for
each line are different. Typical noises are in the range 0.1 K to 0.5 K for
these lines.
A 3-$\sigma$ criterion is used to determine whether a source is detected in a line
or not. But for a few sources, weaker criteria were used.
Detection rate for each line is calculated correspondingly.
Table \ref{tab:brief_sum} is a brief summary of our survey. 
Apparently the $^{12}$CO lines usually have more velocity components detectable than $^{13}$CO and C$^{18}$O,
even if in the same bandwidth. The number of sources with just one component in $^{13}$CO is less than that
of $^{12}$CO and C$^{18}$O, because for weak sources $^{13}$CO may be undetectable on one hand, 
and for strong sources $^{13}$CO might have more than one components on the other hand.

\clearpage
\begin{table}
\caption{Summary of the survey.}
\label{tab:brief_sum}
\begin{tabular}{l c c c }
\hline\hline
$^\#$Components & $^{12}$CO & $^{13}$CO & C$^{18}$O \\
\hline
None   & 6  (10\%) & 18 (30\%) & 23 (38\%) \\
One    & 19 (31\%) & 16 (26\%) & 26 (42\%) \\
Two    & 11 (18\%) & 16 (26\%) & 8 (13\%)\\
Three  & 14 (23\%) & 8 (13\%) & 3 (5\%) \\
Four   & 5  (8\%) &  3 (5\%) & 1 (2\%) \\
More   & 6  (10\%) &  0 & 0 \\
\hline
Detected &  55   &   43   & 38  \\
Total & 61 & 61 & 61 \\
Detection Rate  &  90\% &   71\% & 62\%\\
\hline
\end{tabular}
\end{table}
\clearpage

For those sources with multiple velocity components, 
it is more likely that the components that are detectable in the rarer isotope lines
or those with the largest optical depths are physically associated with the infrared extinction features,
although the ultimate way to determine the velocity corresponding to the extinction feature is to fully map the
region, which is rather time consuming and yet to be carried out.
In practice, we used the following criteria: if a source is detected in C$^{18}$O, then
the components of the three CO isotope lines corresponding to those detected in C$^{18}$O
are recorded (with precedence level assigned according to their optical depths);
if it is not detected in C$^{18}$O, then we use $^{13}$CO to determine which
components to use; if it is neither detected in $^{13}$CO, then all the components of
$^{12}$CO are recorded.

In Table \ref{tab:results_61src_obs} we list the observational
results of our single point survey. The source names used during the observation are just simple
combinations of the Galactic coordinates, which are in accordance with the Simon catalog.
In this table, sources with multiple components have more than one records labeled
with number followed by letter a, b, c, \ldots, in descending order of C$^{18}$O optical depth if it is detected,
otherwise $^{13}$CO optical depth or $^{12}$CO intensity is used for ordering.
For each cloud, the LSR velocity is taken to be the Gaussian fit velocity of $^{13}$CO\ if
$^{13}$CO\ is detected, otherwise that of $^{12}$CO\ is
used. This is because $^{12}$CO lines tend to be saturated and the Gaussian fit
velocity may be inaccurate, while C$^{18}$O lines suffer from the low signal to noise ratio.
As 15 sources have already been mapped by \citet{Simon2006b} in $^{13}$CO,
we compared our results with theirs, and find that the velocities we adopted
are generally in agreement with theirs, with just one exception, viz., G050.39-00.41.
For this source, the velocity component at V$_{LSR}$=41 km s$^{-1}$ is evidently self-absorbed in $^{12}$CO, 
therefore the optical depth of $^{13}$CO and C$^{18}$O of this component could be overestimated.
Thus for this source we take the velocity of its first component to be in accordance with \citet{Simon2006b}.
G028.51+03.60, G038.77+00.78, and G084.81-01.09 also have self-absorption features in $^{12}$CO,
but each of them have only one component in C$^{18}$O, so there's no confusion for these three sources.
This comparison might indicate a reliablity of 94\% of the LSR velocities allocated to each source.

In Table \ref{tab:results_61src_phy} we list the physical properties of these sources.
Parameters derived from $^{13}$CO\ and C$^{18}$O\ are both listed here for comparison.
For sources with multiple components, the first components in Table \ref{tab:results_61src_obs}
are used to derive the physical parameters.
The excitation temperatures are derived from the peak $^{12}$CO\ temperature; for sources with apparent self-absorption,
the values derived this way may be unreliable.
Distance of each source is derived from the LSR velocity using a Galactic rotation curve model given by \citet{Fich1989a}.
The adopted distances are always the nearer one, as the IRDCs appear to be extinction features.
We also tried the rotation curve of \citet{Clemens1985a}, 
and found that the differences in the results of these two models are typically less than 0.2 kpc, with an upper limit of 0.5 kpc.
Following \citet{Simon2006b}, the source sizes are calculated using the 
kinematic distances derived here and the major and minor axis lengths in the Simon catalog.
For sources in the outer galaxy, their angular sizes are estimated from the MSX images.

The column densities of H$_2$ are derived assuming a simple LTE model,
with a moderate $^{13}$CO\ abundance of $1\times10^{-6}$,
and C$^{18}$O\ abundance of $1.4\times10^{-7}$ (\citet{Dickman1978a, Burgh2007a, Kramer1999a, Frerking1982a}).
The volume densities are derived by simply dividing the column densities by the sizes,
and the masses are just the products of column densities and areas.
Although the volume densities and masses derived here are not very reliable,
as the sizes of the extinction features calculated here may not reflect the actual molecular cloud sizes,
they should be qualitatively correct.
In comparison with the results of \citet{Simon2006b}, we find that our results are consistent with theirs within a factor of 2.

\clearpage
\pagestyle{empty}
\setlength{\voffset}{25mm}
\begin{deluxetable}{l c r r r r c c r r r r c c r r r r c}\tablecolumns{19} \tablewidth{0pc} \rotate \tabletypesize{\scriptsize}\tablecaption{Observed properties of the IRDCs.\label{tab:results_61src_obs}}\tablehead{\colhead{} & \colhead{} & \multicolumn{5}{c}{$^{12}$CO} & & \multicolumn{5}{c}{$^{13}$CO} & & \multicolumn{5}{c}{C$^{18}$O} \\\cline{3-7}  \cline{9-13}  \cline{15-19}\\\colhead{} & \colhead{Name} & \colhead{$T_R^\ast$} & \colhead{$\int T_R^\ast d\rm V$} & \colhead{V$_{\rm LSR}$}  & \colhead{$\Delta$V} & \colhead{$\sigma$} & \colhead{} & \colhead{$T_R^\ast$} & \colhead{$\int T_R^\ast d\rm V$} & \colhead{V$_{\rm LSR}$}  & \colhead{$\Delta$V} & \colhead{$\sigma$} & \colhead{} & \colhead{$T_R^\ast$} & \colhead{$\int T_R^\ast d\rm V$} & \colhead{V$_{\rm LSR}$}  & \colhead{$\Delta$V} & \colhead{$\sigma$}\\\colhead{} & \colhead{} & \colhead{(K)} &  \colhead{(K km s$^{-1}$)}  & \colhead{(km s$^{-1}$)} & \colhead{(km s$^{-1}$)}  & \colhead{(K)} & \colhead{} & \colhead{(K)} &  \colhead{(K km s$^{-1}$)}  & \colhead{(km s$^{-1}$)} & \colhead{(km s$^{-1}$)}  & \colhead{(K)} & \colhead{} & \colhead{(K)} &  \colhead{(K km s$^{-1}$)}  & \colhead{(km s$^{-1}$)} & \colhead{(km s$^{-1}$)}  & \colhead{(K)}\\\colhead{(1)}     & \colhead{(2)}   &  \colhead{(3)}   &  \colhead{(4)} &  \colhead{(5)}  &   \colhead{(6)} & \colhead{(7)} & \colhead{} & \colhead{(8)}  & \colhead{(9)} & \colhead{(10)}  & \colhead{(11)}  & \colhead{(12)} & \colhead{} & \colhead{(13)}  & \colhead{(14)}  & \colhead{(15)}  & \colhead{(16)}  & \colhead{(17)}}\startdata
 1a  &G008.21+00.47  &   6.74  &  21.72  &  11.47  &   3.03  &   0.35  &   &   6.20  &  16.11  &  12.02  &   2.44  &   0.50  &   &   1.48  &   3.30  &  11.55  &   2.10  &   0.28 \\
 1b  &   ---     &   5.98  &  22.63  &  14.96  &   3.56  &   ---     &   &   4.35  &   8.87  &  15.02  &   1.91  &   ---     &   &   0.87  &   1.28  &  14.71  &   1.38  &   ---    \\
 2  &G008.67$-$00.70  &  10.26  &  91.11  &  17.61  &   8.34  &   0.19  &   &   6.81  &  28.63  &  17.84  &   3.94  &   0.17  &   &   1.22  &   4.69  &  17.87  &   3.61  &   0.13 \\
 3  &G010.71$-$00.16  &   6.54  & 102.59  &  30.19  &  14.74  &   0.52  &   &   4.74  &  35.94  &  30.06  &   7.13  &   0.57  &   &   1.37  &   8.96  &  30.32  &   6.11  &   0.46 \\
 4  &G010.99$-$00.07  &   7.06  &  62.76  &  29.37  &   8.37  &   0.15  &   &   6.20  &  26.00  &  29.39  &   3.93  &   0.15  &   &   2.35  &   6.61  &  29.38  &   2.64  &   0.09 \\
 5  &G011.87$-$00.62  &   8.72  & 114.33  &  35.01  &  12.31  &   0.22  &   &   3.94  &  21.04  &  35.95  &   5.01  &   0.20  &   &   1.59  &   6.06  &  36.49  &   3.55  &   0.20 \\
 6a  &G012.81+00.36  &   9.50  &  48.26  &  19.05  &   4.78  &   0.17  &   &   7.91  &  23.46  &  18.92  &   2.78  &   0.20  &   &   1.74  &   4.11  &  18.86  &   2.22  &   0.17 \\
 6b  &   ---     &   6.15  &  27.59  &  30.08  &   4.21  &   ---     &   &   3.78  &   8.22  &  30.17  &   2.05  &   ---     &   &   0.94  &   1.11  &  30.06  &   1.10  &   ---    \\
 7a  &G013.22$-$00.07  &   8.69  &  66.69  &  36.24  &   7.22  &   0.52  &   &   8.13  &  30.26  &  36.70  &   3.49  &   0.46  &   &   2.02  &   5.80  &  36.93  &   2.70  &   0.48 \\
 7b  &   ---     &   8.61  &  70.41  &  53.12  &   7.68  &   ---     &   &   6.26  &  42.19  &  52.98  &   6.33  &   ---     &   &   1.41  &   7.26  &  52.92  &   4.86  &   ---    \\
 7c  &   ---     &   3.91  &  13.87  &  43.63  &   3.34  &   ---     &   &   1.63  &  10.07  &  43.14  &   5.81  &   ---     &   &   1.20  &   1.33  &  44.58  &   1.04  &   ---    \\
 8  &G013.34+00.17  &   9.83  &  71.57  &  18.49  &   6.84  &   0.13  &   &   7.31  &  21.54  &  18.46  &   2.77  &   0.17  &   &   1.20  &   2.69  &  18.51  &   2.09  &   0.11 \\
 9a  &G013.82$-$00.48  &   8.59  & 107.43  &  21.24  &  11.75  &   0.78  &   &   5.48  &  38.15  &  21.92  &   6.54  &   0.17  &   &   1.76  &   4.33  &  22.53  &   2.31  &   0.11 \\
 9b  &   ---     &  10.00  &  44.44  &  39.43  &   4.17  &   ---     &   &   5.37  &  13.81  &  39.51  &   2.42  &   ---     &   &   0.83  &   1.24  &  39.56  &   1.38  &   ---    \\
10  &G013.97$-$00.43  &  17.13  & 169.30  &  20.25  &   9.28  &   0.15  &   &  10.26  &  62.43  &  21.45  &   5.71  &   0.15  &   &   3.09  &   6.20  &  22.49  &   1.89  &   0.15 \\
11  &G014.29$-$00.66  &  11.13  &  91.70  &  20.62  &   7.74  &   0.13  &   &   9.11  &  41.98  &  20.74  &   4.33  &   0.11  &   &   2.89  &   8.46  &  21.04  &   2.75  &   0.11 \\
12  &G014.56$-$00.78  &   9.59  &  59.31  &  20.47  &   5.81  &   0.09  &   &   8.72  &  24.63  &  20.63  &   2.65  &   0.09  &   &   2.20  &   4.44  &  20.54  &   1.90  &   0.06 \\
13  &G014.72$-$00.88  &  10.37  &  42.22  &  19.35  &   3.83  &   0.07  &   &   7.41  &  18.91  &  19.13  &   2.40  &   0.13  &   &   0.93  &   1.57  &  19.06  &   1.60  &   0.11 \\
14  &G014.97+01.60  &\nodata  &\nodata  &\nodata  &\nodata  &   0.31  &   &\nodata  &\nodata  &\nodata  &\nodata  &   0.41  &   &\nodata  &\nodata  &\nodata  &\nodata  &   0.41 \\
15a  &G015.80$-$00.40  &   5.91  &  11.61  &  46.91  &   1.85  &   0.56  &   &   3.13  &   5.63  &  46.87  &   1.70  &   0.54  &   &\nodata  &\nodata  &\nodata  &\nodata  &   0.43 \\
15b  &   ---     &   3.80  &  18.48  &  52.09  &   4.58  &   ---     &   &   1.59  &   3.89  &  51.74  &   2.28  &   ---     &   &\nodata  &\nodata  &\nodata  &\nodata  &   ---    \\
16  &G017.09+00.45  &  13.96  &  79.98  &  22.82  &   5.38  &   0.28  &   &   9.81  &  28.00  &  23.45  &   2.68  &   0.26  &   &   2.91  &   5.28  &  23.58  &   1.71  &   0.20 \\
17  &G017.98+01.97  &\nodata  &\nodata  &\nodata  &\nodata  &   0.35  &   &\nodata  &\nodata  &\nodata  &\nodata  &   0.44  &   &\nodata  &\nodata  &\nodata  &\nodata  &   0.37 \\
18  &G019.27+00.07  &   9.69  &  64.78  &  26.31  &   6.29  &   0.26  &   &   7.41  &  29.96  &  26.63  &   3.80  &   0.24  &   &   2.24  &   7.30  &  26.72  &   3.05  &   0.15 \\
19  &G019.92$-$00.29  &   6.33  &  48.56  &  69.15  &   7.20  &   0.56  &   &   3.67  &  20.24  &  69.32  &   5.18  &   0.52  &   &   1.31  &   5.26  &  70.11  &   3.73  &   0.46 \\
20a  &G022.35+00.41  &   5.61  &  24.72  &  60.80  &   4.13  &   0.41  &   &   5.06  &  10.96  &  60.54  &   2.03  &   0.43  &   &   2.35  &   2.59  &  60.59  &   1.03  &   0.43 \\
20b  &   ---     &  10.37  &  55.61  &  84.64  &   5.04  &   ---     &   &   6.87  &  22.78  &  84.22  &   3.11  &   ---     &   &   2.09  &   6.48  &  84.23  &   2.91  &   ---    \\
20c  &   ---     &   6.80  &  23.78  &  53.08  &   3.29  &   ---     &   &   4.46  &   9.57  &  53.16  &   2.01  &   ---     &   &   1.44  &   1.74  &  53.23  &   1.14  &   ---    \\
21a  &G023.42$-$00.52  &   5.89  &  39.89  &  61.68  &   6.36  &   0.41  &   &   4.76  &  14.81  &  62.46  &   2.92  &   0.41  &   &   2.30  &   4.63  &  62.51  &   1.89  &   0.39 \\
21b  &   ---     &   3.30  &  13.76  &  67.91  &   3.93  &   ---     &   &   2.09  &   6.46  &  67.65  &   2.89  &   ---     &   &   0.57  &   1.46  &  67.63  &   2.43  &   ---    \\
22a  &G024.49$-$00.69  &   8.30  &  46.91  &  48.37  &   5.31  &   0.17  &   &   6.06  &  18.70  &  48.44  &   2.91  &   0.15  &   &   1.63  &   4.30  &  48.73  &   2.49  &   0.13 \\
22b  &   ---     &   4.15  &  38.59  &  58.27  &   8.76  &   ---     &   &   1.44  &   9.20  &  58.00  &   5.99  &   ---     &   &\nodata  &\nodata  &\nodata  &\nodata  &   ---    \\
23  &G025.04$-$00.20  &  10.67  &  77.98  &  46.89  &   6.87  &   0.19  &   &   6.46  &  34.13  &  46.71  &   4.95  &   0.09  &   &   2.04  &   5.20  &  47.19  &   2.39  &   0.09 \\
24  &G028.23$-$00.19  &   8.15  &  92.00  &  78.24  &  10.62  &   0.31  &   &   5.02  &  36.61  &  77.77  &   6.84  &   0.26  &   &   1.43  &   7.59  &  77.77  &   5.01  &   0.24 \\
25a  &G028.37+00.07  &   8.69  & 113.24  &  78.73  &  12.25  &   0.44  &   &   6.37  &  49.44  &  78.28  &   7.28  &   0.46  &   &   1.80  &  11.15  &  78.92  &   5.81  &   0.41 \\
25b  &   ---     &   4.11  &  63.63  & 100.82  &  14.57  &   ---     &   &   2.59  &   9.22  & 101.29  &   3.33  &   ---     &   &   1.59  &   2.30  & 101.06  &   1.35  &   ---    \\
26  &G028.51+03.60  &   7.11  &  19.54  &   6.76  &   5.43  &   0.17  &   &   3.57  &  15.69  &   6.47  &   4.13  &   0.19  &   &   2.04  &   3.52  &   7.23  &   1.62  &   0.13 \\
27a  &G028.67+00.13  &   4.31  &  61.91  &  78.19  &  13.50  &   0.30  &   &   3.98  &  23.61  &  78.52  &   5.57  &   0.30  &   &   0.87  &   5.46  &  78.92  &   5.28  &   0.26 \\
27b  &   ---     &   5.26  &  41.67  &  96.96  &   7.45  &   ---     &   &   1.65  &  10.54  &  97.44  &   5.99  &   ---     &   &   0.41  &   1.85  &  98.83  &   4.32  &   ---    \\
28  &G031.97+00.07  &   9.63  &  82.69  &  96.02  &   8.07  &   0.39  &   &   5.20  &  29.56  &  95.71  &   5.33  &   0.30  &   &   1.09  &   5.56  &  95.84  &   4.79  &   0.30 \\
29  &G034.24$-$01.25  &   5.54  &  17.41  &  13.58  &   2.96  &   0.13  &   &   5.22  &   7.76  &  13.83  &   1.40  &   0.07  &   &   2.28  &   2.24  &  13.85  &   0.92  &   0.04 \\
30  &G034.77$-$00.55  &   8.63  & 101.94  &  44.24  &  11.10  &   0.35  &   &   6.28  &  48.80  &  43.79  &   7.30  &   0.33  &   &   1.19  &   5.93  &  44.01  &   4.67  &   0.24 \\
31  &G034.77$-$00.55  &   8.63  & 100.89  &  44.59  &  10.97  &   0.15  &   &   5.89  &  42.13  &  43.92  &   6.73  &   0.13  &   &   1.20  &   5.70  &  43.46  &   4.49  &   0.11 \\
32  &G035.19$-$00.72  &  12.11  &  95.81  &  33.51  &   7.43  &   0.31  &   &   8.43  &  35.00  &  33.34  &   3.90  &   0.24  &   &   1.59  &   4.78  &  33.28  &   2.81  &   0.22 \\
33a  &G035.39$-$00.33  &   9.07  &  51.46  &  44.47  &   5.33  &   0.22  &   &   5.91  &  20.41  &  44.77  &   3.25  &   0.39  &   &   1.41  &   2.56  &  44.73  &   1.70  &   0.30 \\
33b  &   ---     &   4.06  &  37.76  &  55.94  &   8.76  &   ---     &   &   2.33  &  12.52  &  55.97  &   5.04  &   ---     &   &   0.80  &   0.78  &  55.15  &   0.91  &   ---    \\
33c  &   ---     &   5.11  &  13.07  &  27.42  &   2.40  &   ---     &   &   2.41  &   5.02  &  27.18  &   1.96  &   ---     &   &   0.50  &   0.83  &  26.66  &   1.57  &   ---    \\
33d  &   ---     &   3.22  &   8.15  &  13.75  &   2.37  &   ---     &   &   1.07  &   2.46  &  13.43  &   2.17  &   ---     &   &   0.65  &   0.67  &  13.26  &   0.99  &   ---    \\
34a  &G036.67$-$00.11  &   4.48  &  35.81  &  54.34  &   7.52  &   0.19  &   &   4.46  &  18.63  &  54.29  &   3.93  &   0.20  &   &   1.89  &   3.80  &  54.07  &   1.89  &   0.26 \\
34b  &   ---     &   2.59  &  38.17  &  62.43  &  13.85  &   ---     &   &   1.61  &  10.06  &  61.57  &   5.89  &   ---     &   &   0.46  &   0.91  &  61.58  &   1.81  &   ---    \\
34c  &   ---     &   4.50  &  50.33  &  81.11  &  10.50  &   ---     &   &   2.43  &  12.96  &  80.30  &   5.02  &   ---     &   &   0.74  &   2.87  &  80.02  &   3.66  &   ---    \\
35  &G038.77+00.78  &   4.06  &  38.78  &  33.07  &   8.98  &   0.24  &   &   3.83  &  20.44  &  32.14  &   5.01  &   0.20  &   &   0.78  &   3.37  &  32.13  &   4.07  &   0.17 \\
36  &G038.95$-$00.47  &  13.48  &  63.91  &  41.93  &   4.45  &   0.15  &   &  10.44  &  30.39  &  41.96  &   2.73  &   0.15  &   &   2.41  &   5.65  &  41.84  &   2.21  &   0.11 \\
37a  &G050.39$-$00.41  &   4.87  &  57.31  &  42.11  &  14.28  &   0.15  &   &   4.22  &  18.80  &  40.93  &   4.19  &   0.13  &   &   2.13  &   4.35  &  40.84  &   1.93  &   0.11 \\
37b  &   ---     &   4.35  &  31.57  &  63.67  &   6.81  &   ---     &   &   2.65  &  11.48  &  64.09  &   4.08  &   ---     &   &   0.93  &   2.04  &  64.62  &   2.05  &   ---    \\
38  &G076.64$-$01.13  &   1.02  &   1.76  &   8.39  &   1.64  &   0.15  &   &\nodata  &\nodata  &\nodata  &\nodata  &   0.20  &   &\nodata  &\nodata  &\nodata  &\nodata  &   0.17 \\
39  &G076.79+02.59  &   0.96  &   2.80  &   2.48  &   2.73  &   0.39  &   &\nodata  &\nodata  &\nodata  &\nodata  &   0.15  &   &\nodata  &\nodata  &\nodata  &\nodata  &   0.17 \\
40  &G077.61+02.10  &\nodata  &\nodata  &\nodata  &\nodata  &   0.20  &   &\nodata  &\nodata  &\nodata  &\nodata  &   0.22  &   &\nodata  &\nodata  &\nodata  &\nodata  &   0.22 \\
41  &G077.95+02.59  &   1.31  &   2.09  &   3.05  &   1.49  &   0.24  &   &\nodata  &\nodata  &\nodata  &\nodata  &   0.19  &   &\nodata  &\nodata  &\nodata  &\nodata  &   0.17 \\
42  &G078.06$-$00.67  &  11.83  &  55.22  &  $-$0.18  &   4.38  &   0.15  &   &   6.09  &  19.19  &  $-$0.07  &   2.95  &   0.07  &   &   0.81  &   2.11  &  $-$0.24  &   2.42  &   0.09 \\
43  &G078.60+03.92  &   0.44  &   3.33  &  $-$2.12  &   7.19  &   0.11  &   &\nodata  &\nodata  &\nodata  &\nodata  &   0.13  &   &\nodata  &\nodata  &\nodata  &\nodata  &   0.11 \\
44a  &G078.62$-$00.93  &   3.15  &   5.19  &   5.53  &   1.55  &   0.22  &   &\nodata  &\nodata  &\nodata  &\nodata  &   0.22  &   &\nodata  &\nodata  &\nodata  &\nodata  &   0.20 \\
44b  &   ---     &   1.28  &   3.61  &  10.63  &   2.67  &   ---     &   &\nodata  &\nodata  &\nodata  &\nodata  &   ---     &   &\nodata  &\nodata  &\nodata  &\nodata  &   ---    \\
45  &G079.24+00.52  &   5.56  &  23.22  &  $-$0.15  &   3.93  &   0.19  &   &   5.22  &  13.31  &   0.13  &   2.39  &   0.19  &   &   2.09  &   3.63  &   0.26  &   1.63  &   0.17 \\
46  &G079.28+03.25  &   1.56  &   6.35  &  $-$1.72  &   3.83  &   0.35  &   &\nodata  &\nodata  &\nodata  &\nodata  &   0.22  &   &\nodata  &\nodata  &\nodata  &\nodata  &   0.22 \\
47  &G079.58+03.59  &   2.80  &  12.63  &  $-$2.81  &   4.25  &   0.30  &   &   0.44  &   1.04  &  $-$3.57  &   2.19  &   0.20  &   &\nodata  &\nodata  &\nodata  &\nodata  &   0.20 \\
48  &G079.60$-$02.49  &   2.33  &   8.72  &  $-$2.30  &   3.51  &   0.11  &   &\nodata  &\nodata  &\nodata  &\nodata  &   0.17  &   &\nodata  &\nodata  &\nodata  &\nodata  &   0.15 \\
49  &G080.00+02.67  &  19.48  &  69.22  &   5.21  &   3.34  &   0.11  &   &  14.15  &  33.33  &   5.00  &   2.21  &   0.11  &   &   3.20  &   6.06  &   4.92  &   1.77  &   0.07 \\
50a  &G081.52+01.60  &   0.98  &  12.89  &   2.45  &  12.43  &   0.11  &   &\nodata  &\nodata  &\nodata  &\nodata  &   0.11  &   &\nodata  &\nodata  &\nodata  &\nodata  &   0.11 \\
50b  &   ---     &   2.28  &   4.93  &  11.28  &   2.02  &   ---     &   &\nodata  &\nodata  &\nodata  &\nodata  &   ---     &   &\nodata  &\nodata  &\nodata  &\nodata  &   ---    \\
51  &G081.69+02.85  &   3.09  &   7.00  &   5.50  &   2.13  &   0.13  &   &\nodata  &\nodata  &\nodata  &\nodata  &   0.24  &   &\nodata  &\nodata  &\nodata  &\nodata  &   0.24 \\
52  &G084.81$-$01.09  &  10.00  &  79.04  &   2.36  &   7.42  &   0.09  &   &   9.06  &  29.74  &   0.98  &   3.08  &   0.11  &   &   3.26  &   7.35  &   0.94  &   2.12  &   0.07 \\
53a  &G093.14+02.71  &   3.93  &  18.74  & $-$13.08  &   4.47  &   0.07  &   &   0.70  &   2.52  & $-$13.05  &   3.33  &   0.09  &   &\nodata  &\nodata  &\nodata  &\nodata  &   0.11 \\
53b  &   ---     &   3.00  &   8.22  &  $-$2.69  &   2.58  &   ---     &   &   0.59  &   0.78  &  $-$2.68  &   1.23  &   ---     &   &\nodata  &\nodata  &\nodata  &\nodata  &   ---    \\
54  &G110.97$-$00.85  &\nodata  &\nodata  &\nodata  &\nodata  &   0.13  &   &\nodata  &\nodata  &\nodata  &\nodata  &   0.17  &   &\nodata  &\nodata  &\nodata  &\nodata  &   0.15 \\
55  &G111.04$-$00.64  &\nodata  &\nodata  &\nodata  &\nodata  &   0.19  &   &\nodata  &\nodata  &\nodata  &\nodata  &   0.22  &   &\nodata  &\nodata  &\nodata  &\nodata  &   0.20 \\
56a  &G133.28+00.21  &   2.74  &   8.39  & $-$50.46  &   2.86  &   0.15  &   &\nodata  &\nodata  &\nodata  &\nodata  &   0.15  &   &\nodata  &\nodata  &\nodata  &\nodata  &   0.13 \\
56b  &   ---     &   2.44  &   4.33  &  $-$3.08  &   1.67  &   ---     &   &\nodata  &\nodata  &\nodata  &\nodata  &   ---     &   &\nodata  &\nodata  &\nodata  &\nodata  &   ---    \\
57  &G173.38+02.57  &   2.22  &   6.89  & $-$18.00  &   2.91  &   0.28  &   &\nodata  &\nodata  &\nodata  &\nodata  &   0.41  &   &\nodata  &\nodata  &\nodata  &\nodata  &   0.35 \\
58  &G189.97+00.45  &   4.57  &  23.43  &   7.16  &   4.81  &   0.24  &   &   1.09  &   2.80  &   7.57  &   2.42  &   0.31  &   &\nodata  &\nodata  &\nodata  &\nodata  &   0.26 \\
59  &G190.12+00.45  &   2.11  &   9.13  &   7.09  &   4.05  &   0.24  &   &\nodata  &\nodata  &\nodata  &\nodata  &   0.33  &   &\nodata  &\nodata  &\nodata  &\nodata  &   0.31 \\
60  &G206.30$-$02.01  &\nodata  &\nodata  &\nodata  &\nodata  &   0.35  &   &\nodata  &\nodata  &\nodata  &\nodata  &   0.35  &   &\nodata  &\nodata  &\nodata  &\nodata  &   0.37 \\
61  &G206.91$-$02.45  &   4.54  &   9.85  &  18.52  &   2.04  &   0.37  &   &   1.72  &   1.46  &  18.07  &   0.80  &   0.37  &   &\nodata  &\nodata  &\nodata  &\nodata  &   0.31 \\
\enddata\tablecomments{Sources with multiple components are marked with numbers appended with letters a, b, \ldots, in descending order of C$^{18}$O optical depth if it is detected, otherwise in descending order of optical depth of $^{13}$CO or intensity of $^{12}$CO.} \end{deluxetable}

\clearpage
\pagestyle{plaintop}
\setlength{\voffset}{0mm}
\begin{deluxetable}{r c r r r c r c r r c r c r r}\tablecolumns{15} \tablewidth{0pc} \rotate \tabletypesize{\scriptsize}\tablecaption{Physical Parameters of the IRDCs.\label{tab:results_61src_phy}}\tablehead{\colhead{} & \colhead{} & \colhead{} & \colhead{} & \colhead{} &  & \multicolumn{4}{c}{$^{13}$CO} &  & \multicolumn{4}{c}{C$^{18}$O} \\\cline{7-10} \cline{12-15} \\\colhead{} & \colhead{Name} & \colhead{T$_{ex}$}  & \colhead{D} & \colhead{Size} & \colhead{} & \colhead{$\tau$} & \colhead{N(H$_2$)} & \colhead{n(H$_2$)}  & \colhead{mass} & \colhead{} & \colhead{$\tau$} & \colhead{N(H$_2$)} & \colhead{n(H$_2$)}  & \colhead{mass} \\\colhead{} & \colhead{} & \colhead{(K)} & \colhead{(kpc)}  & \colhead{(pc)} & \colhead{} & \colhead{} &  \colhead{($10^{22}$ cm$^{-2}$)}  & \colhead{(cm$^{-3}$)} & \colhead{(M$_\odot$)} & \colhead{} & \colhead{} &  \colhead{($10^{22}$ cm$^{-2}$)}  & \colhead{(cm$^{-3}$)} & \colhead{(M$_\odot$)} \\ \colhead{(1)}     & \colhead{(2)} &  \colhead{(3)}  &   \colhead{(4)} & \colhead{(5)} & \colhead{} & \colhead{(6)}  & \colhead{(7)} & \colhead{(8)}  & \colhead{(9)} & \colhead{} & \colhead{(10)} & \colhead{(11)}  & \colhead{(12)}  & \colhead{(13)} }\startdata 
 1  &  G008.21+00.47  &    10.08  &     2.42  &     7.13  &   &     2.47  & 
   1.24  &  565  &  7363  &   &     0.25  &     1.77  &  805  &  10489  \\ 
 2  &  G008.67$-$00.70  &    13.67  &     3.05  &     7.96  &   &     1.08  & 
   2.51  &  1021  &  18508  &   &     0.13  &     2.86  &  1164  &  21097  \\ 
 3  &  G010.71$-$00.16  &     9.87  &     3.73  &    12.32  &   &     1.28  & 
   2.76  &  725  &  48743  &   &     0.23  &     4.79  &  1259  &  84628  \\ 
 4  &  G010.99$-$00.07  &    10.41  &     3.63  &     7.47  &   &     2.08  & 
   2.03  &  879  &  13174  &   &     0.40  &     3.59  &  1558  &  23325  \\ 
 5  &  G011.87$-$00.62  &    12.11  &     3.92  &     1.42  &   &     0.60  & 
   1.74  &  3984  &  406  &   &     0.20  &     3.49  &  7987  &  814  \\ 
 6  &  G012.81+00.36  &    12.90  &     2.49  &     4.60  &   &     1.76  & 
   2.00  &  1408  &  4916  &   &     0.20  &     2.44  &  1719  &  6000  \\ 
 7  &  G013.22$-$00.07  &    12.07  &     3.77  &     6.20  &   &     2.68  & 
   2.50  &  1306  &  11200  &   &     0.26  &     3.33  &  1742  &  14941  \\ 
 8  &  G013.34+00.17  &    13.23  &     2.39  &     4.28  &   &     1.35  & 
   1.86  &  1406  &  3959  &   &     0.13  &     1.61  &  1221  &  3438  \\ 
 9  &  G013.82$-$00.48  &    11.97  &     2.64  &     2.22  &   &     1.01  & 
   3.14  &  4574  &  1810  &   &     0.23  &     2.48  &  3619  &  1432  \\ 
10  &  G013.97$-$00.43  &    20.60  &     2.58  &     3.54  &   &     0.91  & 
   7.01  &  6417  &  10242  &   &     0.20  &     4.85  &  4443  &  7092  \\ 
11  &  G014.29$-$00.66  &    14.55  &     2.49  &     4.47  &   &     1.69  & 
   3.80  &  2755  &  8858  &   &     0.30  &     5.34  &  3869  &  12438  \\ 
12  &  G014.56$-$00.78  &    12.99  &     2.45  &     3.93  &   &     2.36  & 
   2.10  &  1735  &  3785  &   &     0.26  &     2.64  &  2181  &  4757  \\ 
13  &  G014.72$-$00.88  &    13.78  &     2.30  &     1.05  &   &     1.24  & 
   1.66  &  5122  &  214  &   &     0.09  &     0.96  &  2970  &  124  \\ 
14  &  G014.97+01.60  &  \nodata  &  \nodata  &  \nodata  &   &  \nodata  & 
\nodata  &  \nodata  &  \nodata  &   &  \nodata  &  \nodata  &  \nodata  & 
\nodata  \\ 
15  &  G015.80$-$00.40  &     9.23  &     4.02  &   17.92  &   &     0.75  & 
   0.42  &  76  &  15886  &   &  \nodata  &  \nodata  &  \nodata  &  \nodata
 \\ 
16  &  G017.09+00.45  &    17.41  &     2.44  &     4.71  &   &     1.21  & 
   2.82  &  1939  &  7271  &   &     0.23  &     3.70  &  2546  &  9548  \\ 
17  &  G017.98+01.97  &  \nodata  &  \nodata  &  \nodata  &   &  \nodata  & 
\nodata  &  \nodata  &  \nodata  &   &  \nodata  &  \nodata  &  \nodata  & 
\nodata  \\ 
18  &  G019.27+00.07  &    13.08  &     2.51  &     2.84  &   &     1.43  & 
   2.57  &  2934  &  2409  &   &     0.26  &     4.36  &  4977  &  4087  \\ 
19  &  G019.92$-$00.29  &     9.66  &     4.67  &     9.43  &   &     0.86  & 
   1.54  &  530  &  16001  &   &     0.23  &     2.79  &  959  &  28948  \\ 
20  &  G022.35+00.41  &     8.92  &     4.15  &     7.37  &   &     2.26  & 
   0.82  &  360  &  5196  &   &     0.54  &     1.35  &  593  &  8555  \\ 
21  &  G023.42$-$00.52  &     9.21  &     4.19  &     5.66  &   &     1.63  & 
   1.12  &  639  &  4163  &   &     0.49  &     2.43  &  1390  &  9058  \\ 
22  &  G024.49$-$00.69  &    11.67  &     3.47  &     5.66  &   &     1.30  & 
   1.52  &  872  &  5678  &   &     0.22  &     2.44  &  1395  &  9083  \\ 
23  &  G025.04$-$00.20  &    14.08  &     3.36  &     2.93  &   &     0.92  & 
   3.04  &  3361  &  3033  &   &     0.21  &     3.23  &  3570  &  3222  \\ 
24  &  G028.23$-$00.19  &    11.52  &     4.77  &    14.70  &   &     0.95  & 
   2.97  &  653  &  74702  &   &     0.19  &     4.28  &  943  &  107889  \\ 
25  &  G028.37+00.07  &    12.07  &     4.80  &     2.45  &   &     1.31  & 
   4.08  &  5406  &  2852  &   &     0.23  &     6.41  &  8489  &  4479  \\ 
26  &  G028.51+03.60  &    10.46  &     0.65  &     0.64  &   &     0.69  & 
   1.23  &  6196  &  58  &   &     0.34  &     1.91  &  9678  &  91  \\ 
27  &  G028.67+00.13  &     7.57  &     4.81  &    10.82  &   &     2.48  & 
   1.73  &  518  &  23664  &   &     0.22  &     2.79  &  835  &  38105  \\ 
28  &  G031.97+00.07  &    13.03  &     6.01  &    15.71  &   &     0.77  & 
   2.53  &  521  &  72744  &   &     0.12  &     3.31  &  682  &  95240  \\ 
29  &  G034.24$-$01.25  &     8.84  &     1.14  &     0.98  &   &     2.77  & 
   0.58  &  1924  &  64  &   &     0.53  &     1.16  &  3868  &  129  \\ 
30  &  G034.77$-$00.55  &    12.01  &     3.01  &     3.18  &   &     1.29  & 
   4.02  &  4099  &  4738  &   &     0.15  &     3.40  &  3467  &  4007  \\ 
31  &  G034.77$-$00.55  &    12.01  &     3.02  &     2.04  &   &     1.14  & 
   3.47  &  5520  &  1681  &   &     0.15  &     3.27  &  5205  &  1585  \\ 
32  &  G035.19$-$00.72  &    15.54  &     2.39  &     0.76  &   &     1.18  & 
   3.29  &  13989  &  222  &   &     0.14  &     3.13  &  13303  &  211  \\ 
33  &  G035.39$-$00.33  &    12.46  &     3.07  &     1.17  &   &     1.04  & 
   1.71  &  4750  &  271  &   &     0.17  &     1.49  &  4143  &  236  \\ 
34  &  G036.67$-$00.11  &     7.75  &     3.64  &     9.24  &   &     4.51  & 
   1.37  &  480  &  13614  &   &     0.54  &     1.94  &  680  &  19308  \\ 
35  &  G038.77+00.78  &     7.30  &     2.32  &     1.02  &   &     2.79  & 
   1.50  &  4756  &  182  &   &     0.21  &     1.72  &  5456  &  209  \\ 
36  &  G038.95$-$00.47  &    16.92  &     2.94  &     4.05  &   &     1.48  & 
   3.00  &  2402  &  5745  &   &     0.20  &     3.89  &  3111  &  7439  \\ 
37  &  G050.39$-$00.41  &     8.15  &     3.57  &     2.17  &   &     1.97  & 
   1.39  &  2067  &  762  &   &     0.57  &     2.23  &  3332  &  1229  \\ 
38  &  G076.64$-$01.13  &     3.99  &     1.96  &     3.69  &   &  \nodata  & 
\nodata  &  \nodata  &  \nodata  &   &  \nodata  &  \nodata  &  \nodata  & 
\nodata  \\ 
39  &  G076.79+02.59  &     3.93  &     1.04  &     4.50  &   &  \nodata  & 
\nodata  &  \nodata  &  \nodata  &   &  \nodata  &  \nodata  &  \nodata  & 
\nodata  \\ 
40  &  G077.61+02.10  &  \nodata  &  \nodata  &  \nodata  &   &  \nodata  & 
\nodata  &  \nodata  &  \nodata  &   &  \nodata  &  \nodata  &  \nodata  & 
\nodata  \\ 
41  &  G077.95+02.59  &     4.34  &     1.77  &     1.22  &   &  \nodata  & 
\nodata  &  \nodata  &  \nodata  &   &  \nodata  &  \nodata  &  \nodata  & 
\nodata  \\ 
42  &  G078.06$-$00.67  &    15.26  &     0.43  &     0.80  &   &     0.72  & 
   1.78  &  7196  &  134  &   &     0.07  &     1.37  &  5516  &  102  \\ 
43  &  G078.60+03.92  &     3.30  &     3.37  &     7.12  &   &  \nodata  & 
\nodata  &  \nodata  &  \nodata  &   &  \nodata  &  \nodata  &  \nodata  & 
\nodata  \\ 
44  &  G078.62$-$00.93  &     6.34  &     1.68  &     3.03  &   &  \nodata  & 
\nodata  &  \nodata  &  \nodata  &   &  \nodata  &  \nodata  &  \nodata  & 
\nodata  \\ 
45  &  G079.24+00.52  &     8.86  &     0.55  &     0.28  &   &     2.72  & 
   0.99  &  11629  &  8  &   &     0.47  &     1.89  &  22069  &  16  \\ 
46  &  G079.28+03.25  &     4.61  &     0.08  &     0.08  &   &  \nodata  & 
\nodata  &  \nodata  &  \nodata  &   &  \nodata  &  \nodata  &  \nodata  & 
\nodata  \\ 
47  &  G079.58+03.59  &     5.97  &     3.37  &     4.24  &   &     0.17  & 
   0.08  &  60  &  165  &   &  \nodata  &  \nodata  &  \nodata  &  \nodata  \\ 
48  &  G079.60$-$02.49  &     5.47  &     3.11  &     3.41  &   &  \nodata  & 
\nodata  &  \nodata  &  \nodata  &   &  \nodata  &  \nodata  &  \nodata  & 
\nodata  \\ 
49  &  G080.00+02.67  &    22.96  &     1.48  &     0.74  &   &     1.29  & 
   4.04  &  17789  &  255  &   &     0.18  &     5.11  &  22521  &  322  \\ 
50  &  G081.52+01.60  &     3.95  &     1.25  &     2.47  &   &  \nodata  & 
\nodata  &  \nodata  &  \nodata  &   &  \nodata  &  \nodata  &  \nodata  & 
\nodata  \\ 
51  &  G081.69+02.85  &     6.29  &     1.23  &     2.56  &   &  \nodata  & 
\nodata  &  \nodata  &  \nodata  &   &  \nodata  &  \nodata  &  \nodata  & 
\nodata  \\ 
52  &  G084.81$-$01.09  &    13.40  &     0.77  &     0.53  &   &     2.32  & 
   2.58  &  15811  &  84  &   &     0.39  &     4.44  &  27224  &  144  \\ 
53  &  G093.14+02.71  &     7.17  &     2.35  &     5.02  &   &     0.20  & 
   0.18  &  119  &  544  &   &  \nodata  &  \nodata  &  \nodata  &  \nodata  \\ 
54  &  G110.97$-$00.85  &  \nodata  &  \nodata  &  \nodata  &   &  \nodata  & 
\nodata  &  \nodata  &  \nodata  &   &  \nodata  &  \nodata  &  \nodata  & 
\nodata  \\ 
55  &  G111.04$-$00.64  &  \nodata  &  \nodata  &  \nodata  &   &  \nodata  & 
\nodata  &  \nodata  &  \nodata  &   &  \nodata  &  \nodata  &  \nodata  & 
\nodata  \\ 
56  &  G133.28+00.21  &     5.91  &     4.68  &    14.93  &   &  \nodata  & 
\nodata  &  \nodata  &  \nodata  &   &  \nodata  &  \nodata  &  \nodata  & 
\nodata  \\ 
57  &  G173.38+02.57  &     5.35  &    19.50  &    59.48  &   &  \nodata  & 
\nodata  &  \nodata  &  \nodata  &   &  \nodata  &  \nodata  &  \nodata  & 
\nodata  \\ 
58  &  G189.97+00.45  &     7.85  &     1.99  &     2.24  &   &     0.27  & 
   0.21  &  297  &  120  &   &  \nodata  &  \nodata  &  \nodata  &  \nodata  \cr
59  &  G190.12+00.45  &     5.23  &     1.79  &     3.13  &   &  \nodata  & 
\nodata  &  \nodata  &  \nodata  &   &  \nodata  &  \nodata  &  \nodata  & 
\nodata  \cr
60  &  G206.30$-$02.01  &  \nodata  &  \nodata  &  \nodata  &   &  \nodata  & 
\nodata  &  \nodata  &  \nodata  &   &  \nodata  &  \nodata  &  \nodata  & 
\nodata  \cr
61  &  G206.91$-$02.45  &     7.81  &     1.92  &     1.12  &   &     0.47  & 
   0.11  &  311  &  15  &   &  \nodata  &  \nodata  &  \nodata  &  \nodata  
\enddata
\end{deluxetable}

\clearpage

\section{Analysis}

\subsection{Detection rate}
Using a 3$\sigma$ criterion, the detection rate is 90\% for $^{12}$CO,  71\% for
$^{13}$CO, and 62\% for C$^{18}$O\ (Table \ref{tab:brief_sum}).
The detection rate is higher in the inner Galactic region than in the outer region.
If restricted to the first Galactic quadrant, then the detection rate becomes 94\%, 77\%, and 73\%, for the three lines, respectively;
while the detection rate for these three lines is 67\%, 33\%, and 0\% in the outer regions.
About 60\% of the sources in our sample have multiple components in $^{12}$CO,
but only 70\% of them are detectable in $^{13}$CO, and one third of them are detectable in C$^{18}$O.
If we treat each velocity component as individual sources, 
then the total number of components in $^{12}$CO is 137, while 62\% of them are detectable in $^{13}$CO,
and 40\% of them are detectable in C$^{18}$O, at an RMS noise level of 0.1 K -- 0.4 K.
As the IRDCs are believed to be dense condensations of molecular gases, 
emissions from dense gas tracers such as C$^{18}$O are expected, thus the components without detectable
counterparts in $^{13}$CO or C$^{18}$O may not be associated with the IRDCs.

\subsection{Distribution in the Galaxy}
\begin{figure}[htbp]
\plotone{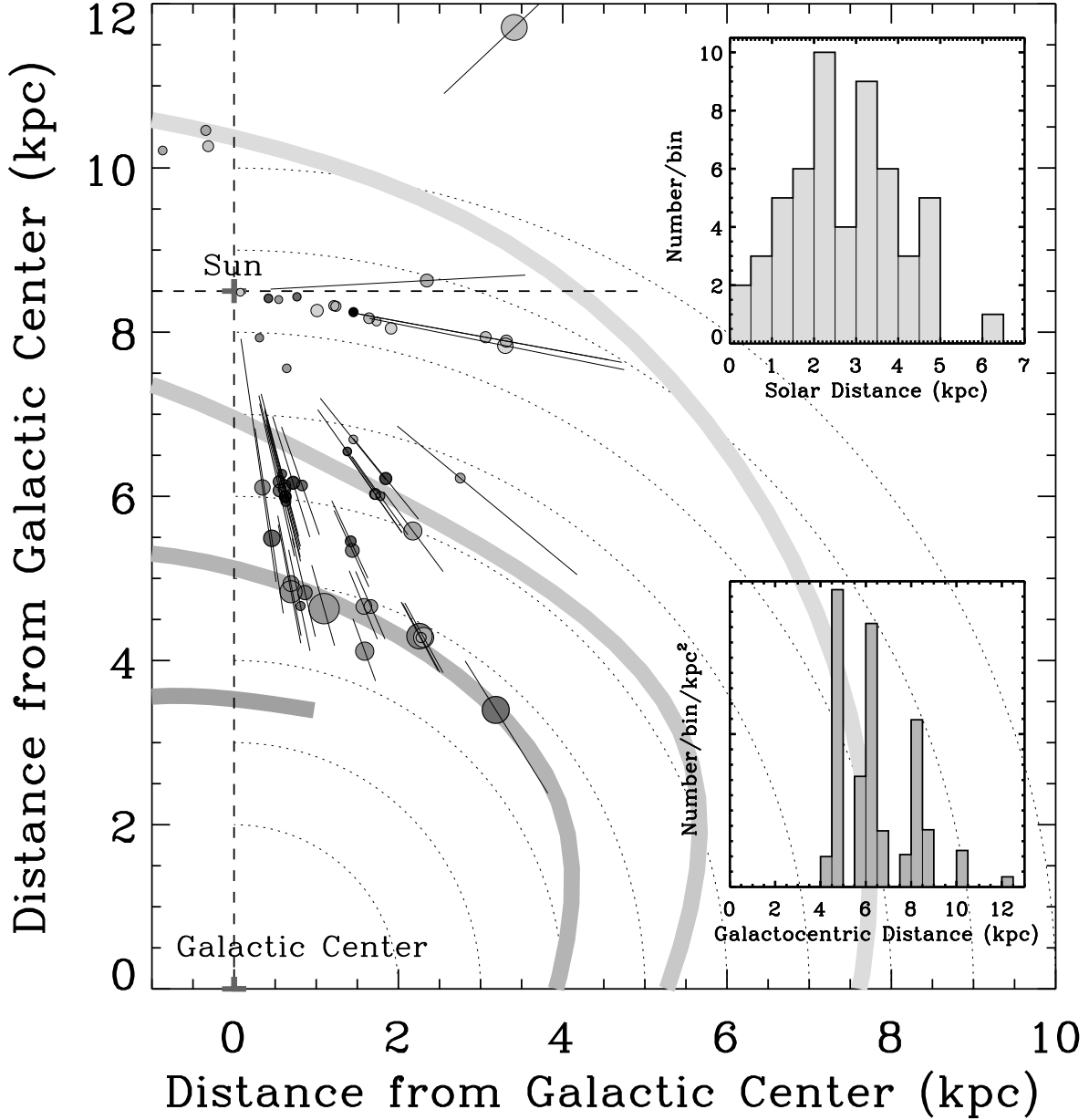}
\figcaption{A face-on view of our sample in the Galactic plane.
Sizes of the circles are roughly proportional to the IRDC sizes, and the grayscales filling the
circles represent the excitation temperatures (the darkest are with the highest T$_{ex}$).
The error bars for sources farther than 2 kpc from the sun represent distance uncertainties calculated using the rotation curve model 
by adding or subtracting their individual V$_{LSR}$ by 10 km s$^{-1}$.
Spiral arms are drawn using cubic splines with fiducial points taken from \citet{Taylor1993a}.
Solar distance distribution is shown in the upper histogram.
The lower panel histogram shows the number distribution with Galactocentric radius,
weighted by the area of the corresponding face-on segment (following \citet{Simon2006b}).
The vertical scale of the histogram is arbitrary, and it is proportional to the surface density of IRDCs in the Galaxy plane.
\label{fig:distri_Gala}}
\end{figure}

The Galactic distribution of our sample is shown in Figure \ref{fig:distri_Gala}, 
overlaid with four spiral arms with parameters from \citet{Taylor1993a} for comparison.
Besides the expected concentration of sources near the sun, 
another obvious feature is that about 60\% of the detected objects are in a ring with Galactocentric distance from $\sim$4 kpc to $\sim$6 kpc,
which is consistent with the 5 kpc molecular ring picture of the Galaxy \citep{Simon2001a}.
Figure \ref{fig:distri_Gala} also shows a slight trace of spiral pattern, 
in that the sources in the first quadrant are seemingly distributed in three segments with about 15 objects in each, 
which looks like the Galactic spiral structure in this region, namely, 
the local spur, the Sagittarius-Carina arm, and the Scuturn-Crux arm \citep{Vallee2005a} (see also \citet{Solomon1989a, Russeil2003a}),
while the Norma arm is missing due to the limited declination coverage of the telescope.
However, uncertainties in the kinematic distances and the small scale of our sample make this pattern insignificant.
Although the errors caused by uncertainties in LSR velocities are not significant, typically less than 0.1 kpc, 
and different rotation curve models usually yield similar results with discrepancies less than 0.5 kpc,
however, random and/or streaming motions of individual clouds can affect distance determination significantly,
which can cause errors as high as several kpc.
Error bars calculated assuming a velocity uncertainty of 10 km s$^{-1}$ are overploted on Figure \ref{fig:distri_Gala} for sources farther than 2 kpc from the sun.
Although we can still see some spiral pattern with these error bars, 
however, irregular velocity caused by streaming motions and shocks can sometimes be as large as 30 km s$^{-1}$ \citep{Brand1993a},
and errors of this scale can entirely damage any spiral pattern.
At this stage it is hard to tell whether the pattern is true or just coincident as our sample is not large and uniform enough.
Actually, if we treat all the velocity components detected in $^{12}$CO as individual sources,
and plot them on the Galactic plane with corresponding kinematic distances, then no sign of spiral pattern can be identified.
Thus we can conjecture that maybe only a subclass of molecular clouds (e.g., the densest and the most massive ones) are distributed in spirals, if there are any spirals at all.

Some sources out of the first quadrant have anomalous LSR velocities, e.g., 
$\rm G173.38+02.57$ in the second quadrant has an LSR velocity of $-18$ km s$^{-1}$,
and the distance derived from the rotation curve is 19.5 kpc, which seems to be unreasonable.
This kind of peculiar velocity is usually attributed to streaming motions \citep{Russeil2007a}
which is believed to be caused by the spiral shocks in the density wave theory.
In Table \ref{tab:results_61src_obs} we notice that the sources in the second quadrant with detectable CO emission
all have rather negative velocities, while those in the third quadrant all have positive velocities,
which is in good agreement with the results of \citet{Russeil2007a} (see also \citet{Brand1993a}),
wherein it is stated that the Perseus arm exhibits minus V$_{LSR}$ departures in the
second quadrant and positive departures in the third quadrant, 
suggesting that these outer sources are in the Perseus arm, rather than further in the Cygnus arm.
The sources not in the first quadrant with loose shapes and lacking detectable $^{13}$CO or C$^{18}$O lines
may be distinct in nature with the other typical IRDCs.
The enrollment of these sources into our sample is merely supplementary.
It is probable that the IRDCs identified in the outer Galaxy are false due to weak and noisy infrared background.

\clearpage
\clearpage

\subsection{Physical properties of the IRDCs}
\begin{figure}[htbp]
\plottwo{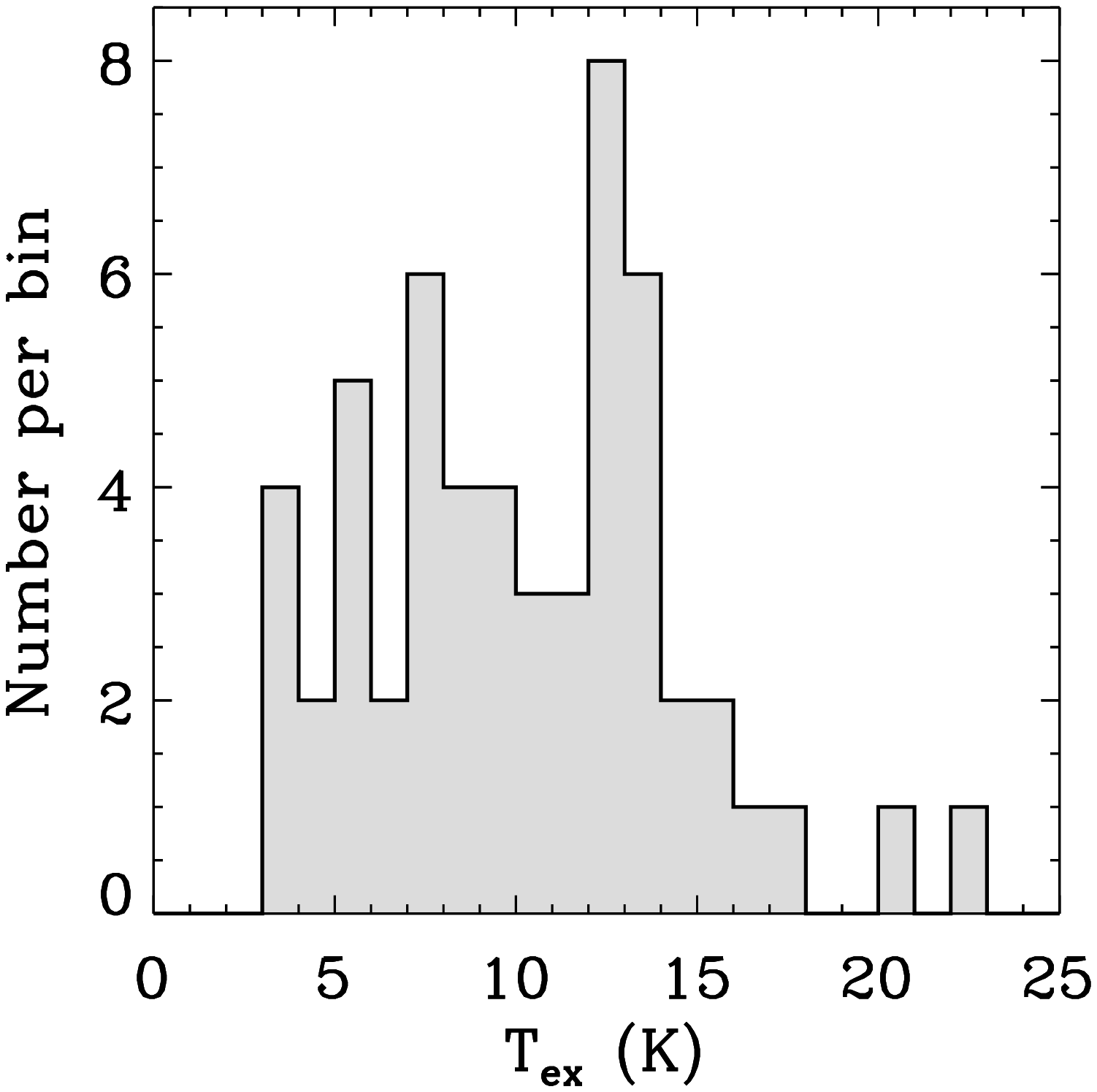}{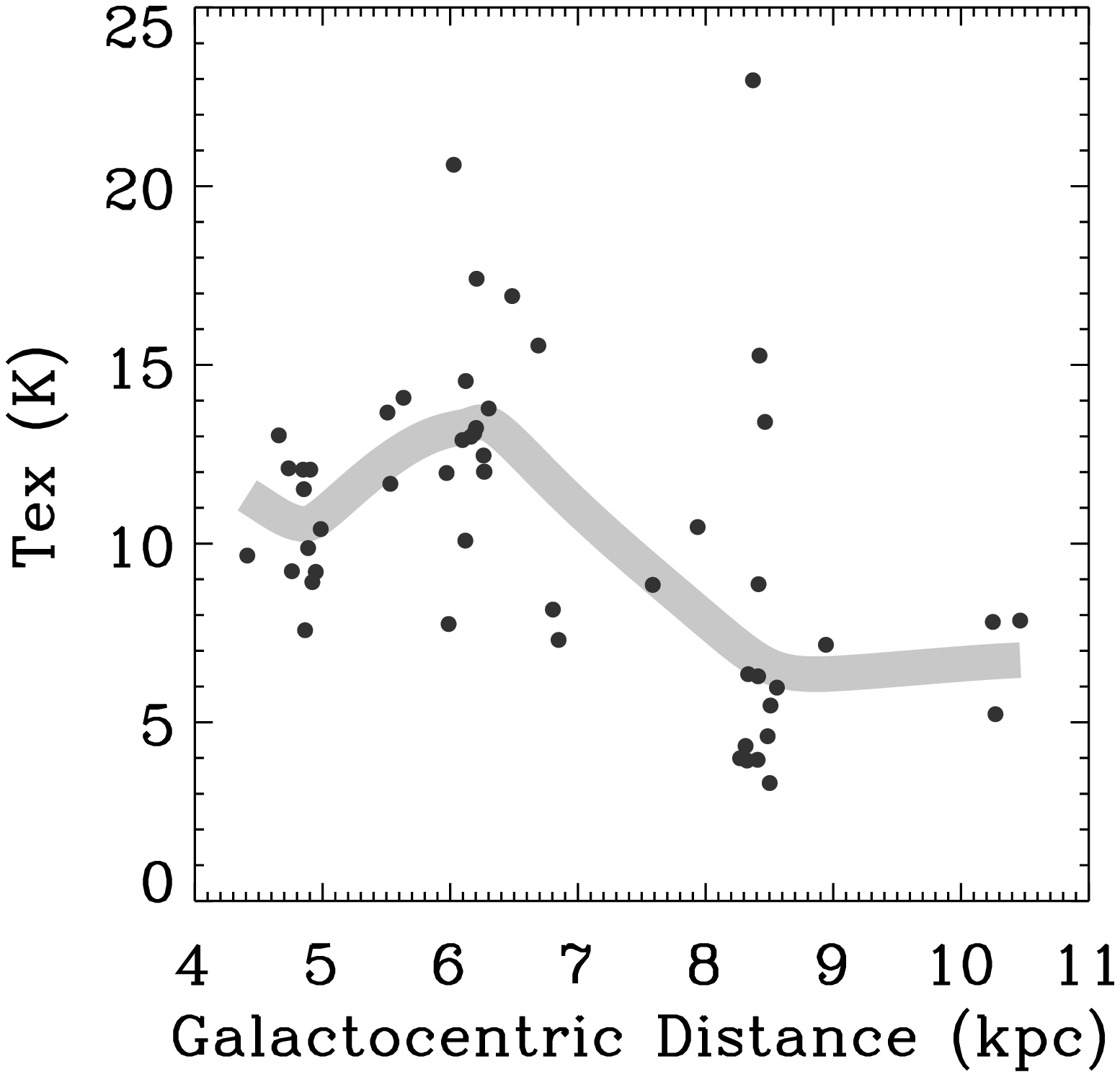}
\figcaption{Histogram of excitation temperature and its variation with Galactocentric distance.
The smooth thick gray curve shows the general trend of the variation, 
with values averaged at each radius (the same for the other similar figures hereinafter, without further notice).
\label{fig:Tex}}
\end{figure}

The mean excitation temperature derived from $^{12}$CO spectrum is 10 K,
which confirms the assumption in \citet{Simon2006b}. 
The mean T$_R^\ast$ are 6 K and 2 K for $^{13}$CO and C$^{18}$O respectively.
A histogram of excitation temperature and its variation with Galactocentric distance is shown in Figure \ref{fig:Tex}.
Excitation temperature is roughly twice higher in the inner Galaxy than in the outer regions.

\begin{figure}[htbp]
\plottwo{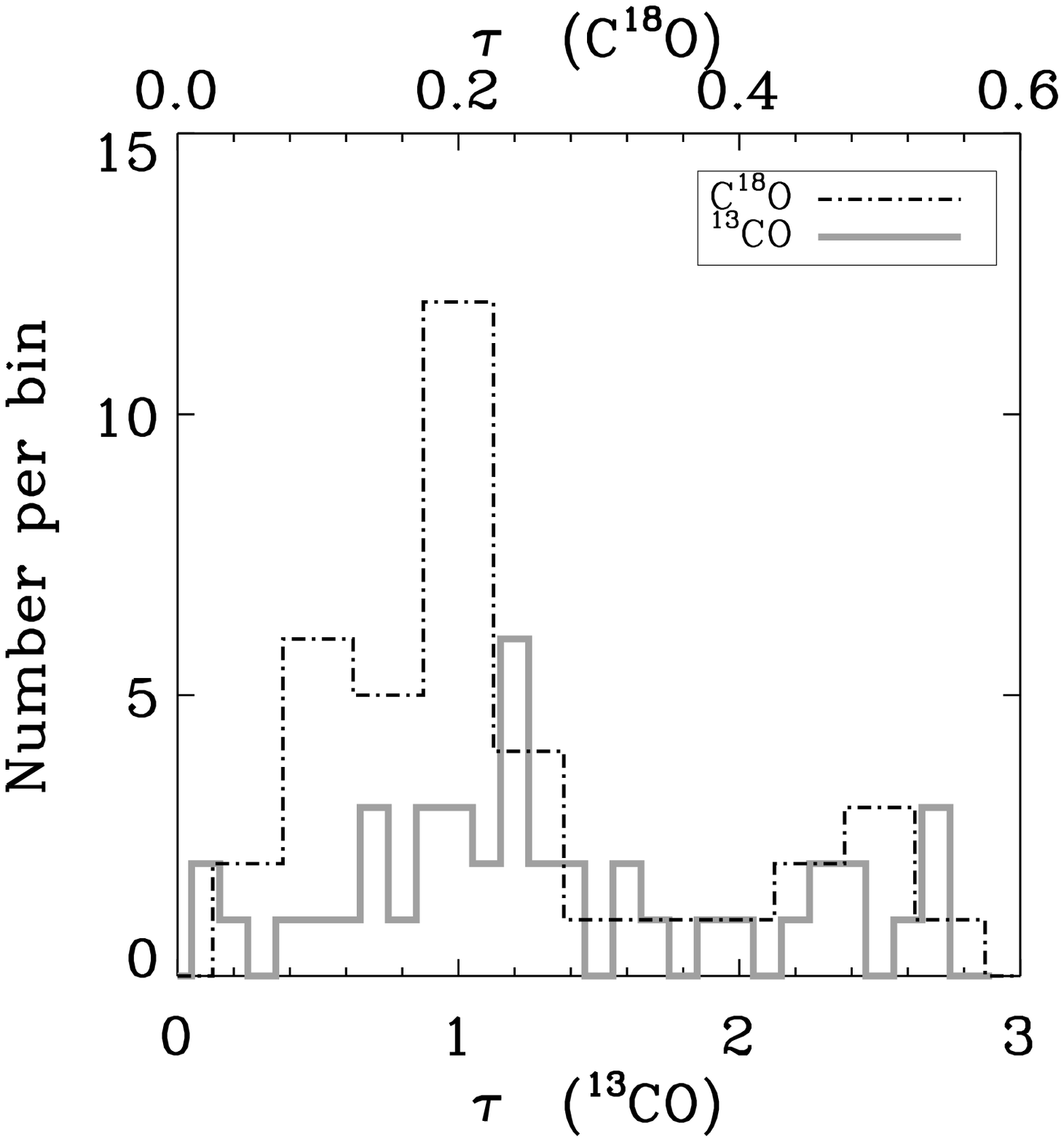}{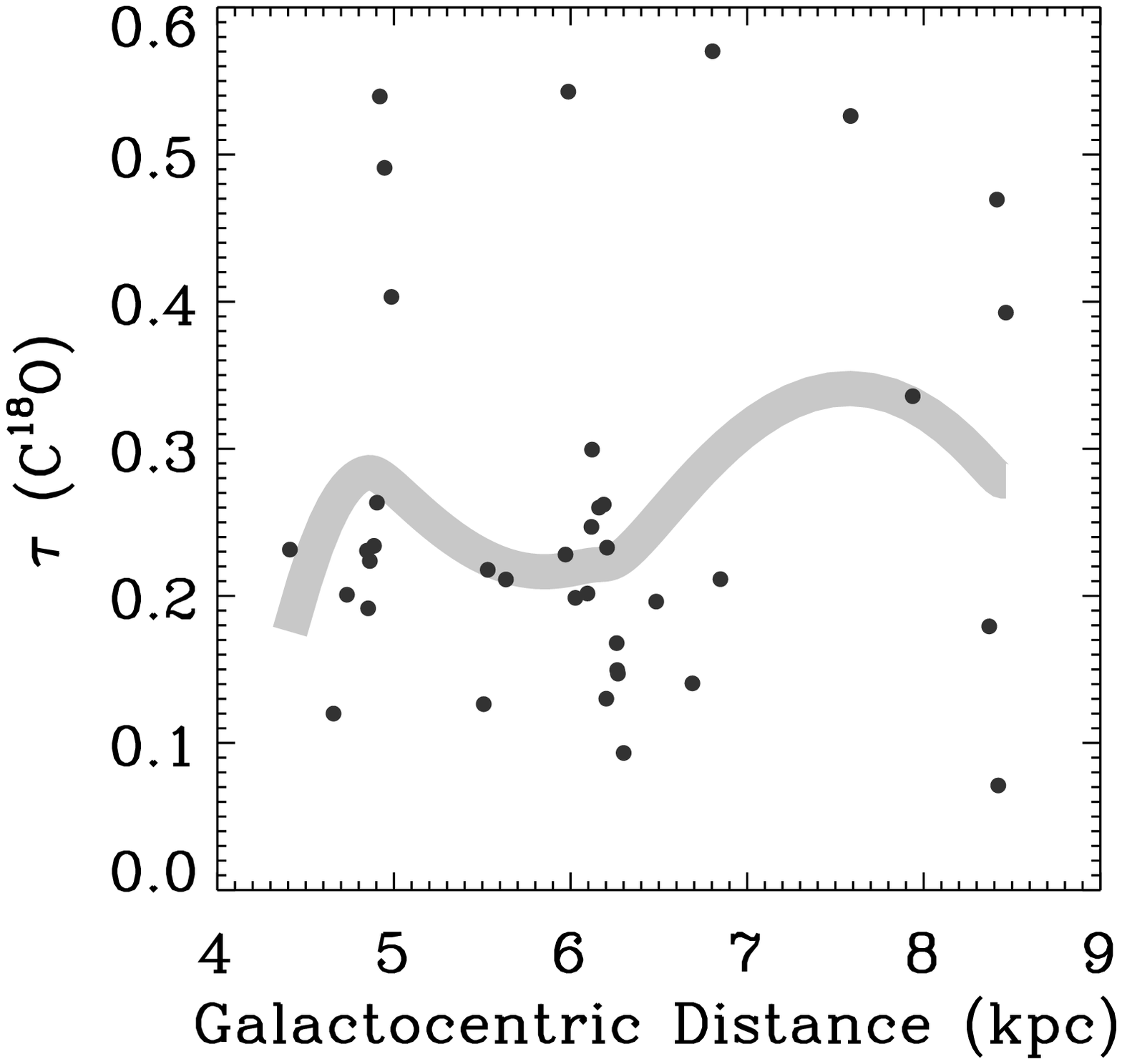}
\figcaption{The left panel shows histograms of the optical depth of $^{13}$CO and C$^{18}$O.
Notice that different horizontal scales are used for $^{13}$CO and C$^{18}$O respectively. 
The right panel shows the variation of the C$^{18}$O optical depth with Galactic radius.
\label{fig:htau}}
\end{figure}

Histogram of optical depths of $^{13}$CO and C$^{18}$O is shown in Figure \ref{fig:htau};
also shown is their variation with Galactocentric distance.
The mean optical depth is 1.4 and 0.3 respectively.
For sources detected in $^{13}$CO, 30\% have optical depth greater than unity,
while almost all sources are thin in C$^{18}$O.
The optical depth ratio of C$^{18}$O to $^{13}$CO has a mean value of $0.16\simeq1/6$, 
which is roughly consistent with the abundance ratio we assumed at the beginning.
Optical depths of the highly saturated lines derived here may be rather inaccurate.

The $^{12}$CO spectrum of several clouds are evidently self-absorbed.
The most apparent ones are G028.51+03.60, G038.77+00.78, and G050.39$-$00.41.
Spectrum of some clouds have indications of wide line-wings, which may be due to outflow motions triggered by star forming activities.

\clearpage
\clearpage

\begin{figure}[htbp]
\plottwo{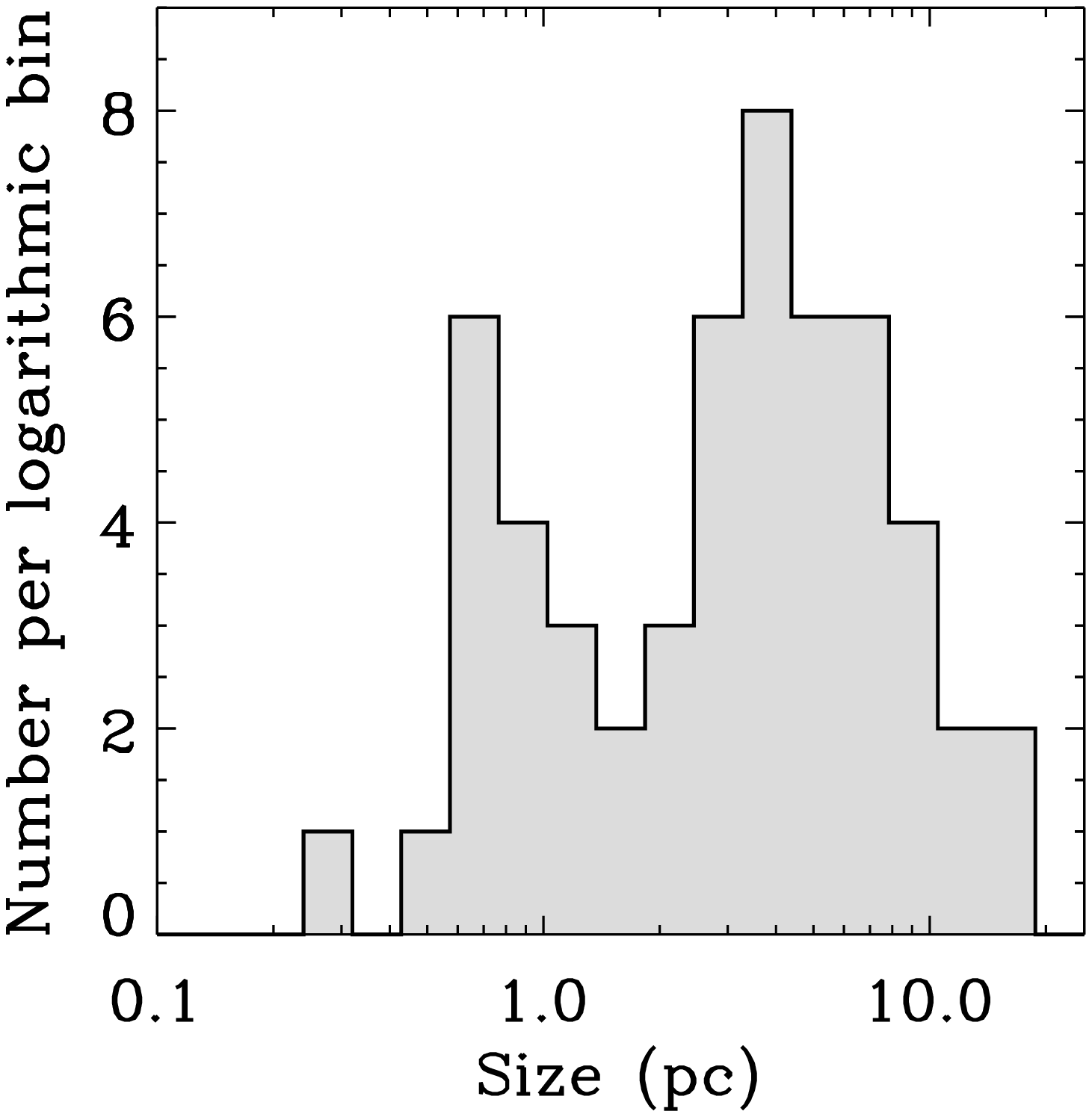}{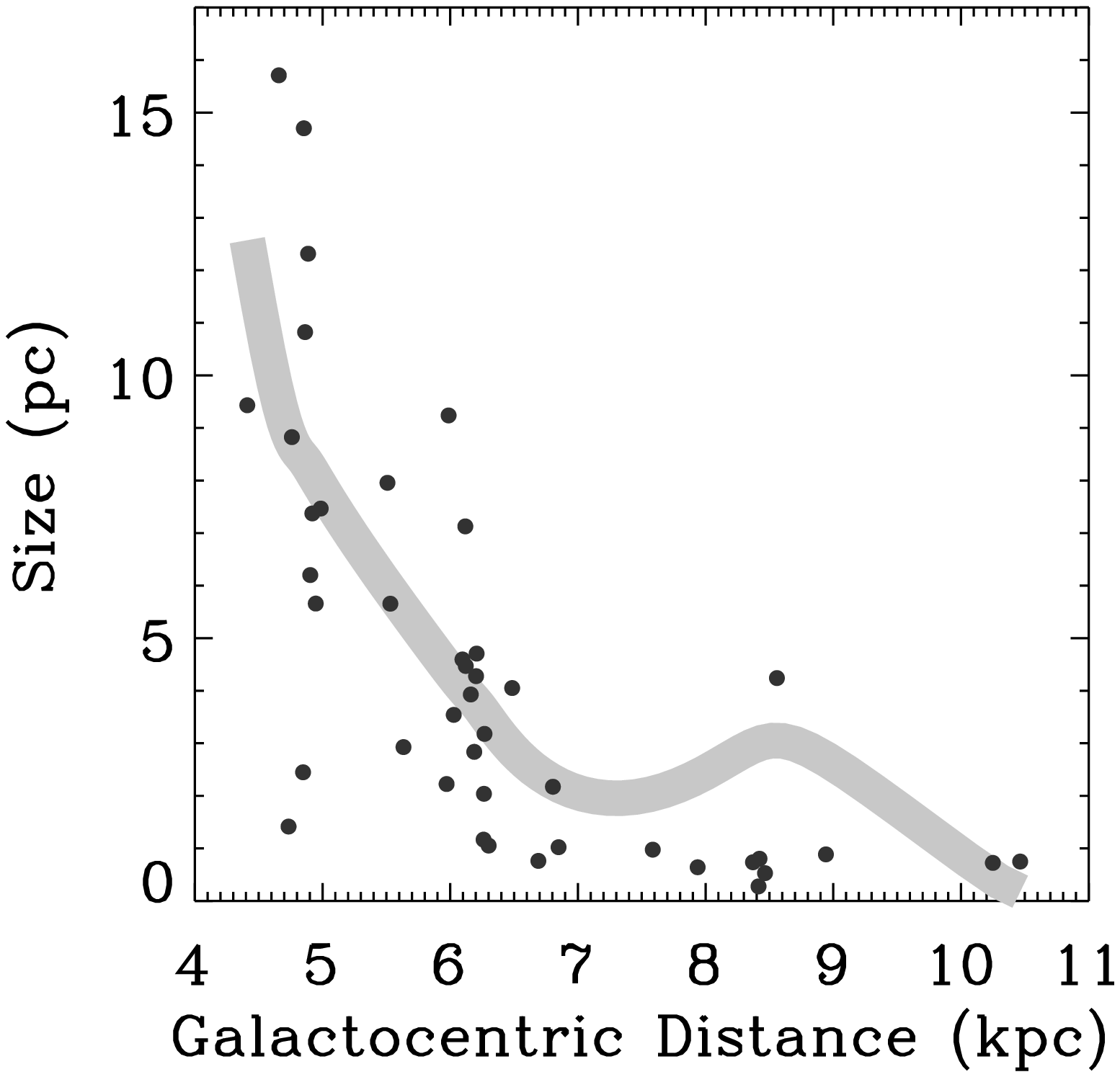}
\figcaption{Histogram of size and its variation with Galactocentric distance. \label{fig:hphysize}}
\end{figure}

\begin{figure}[htbp]
\plottwo{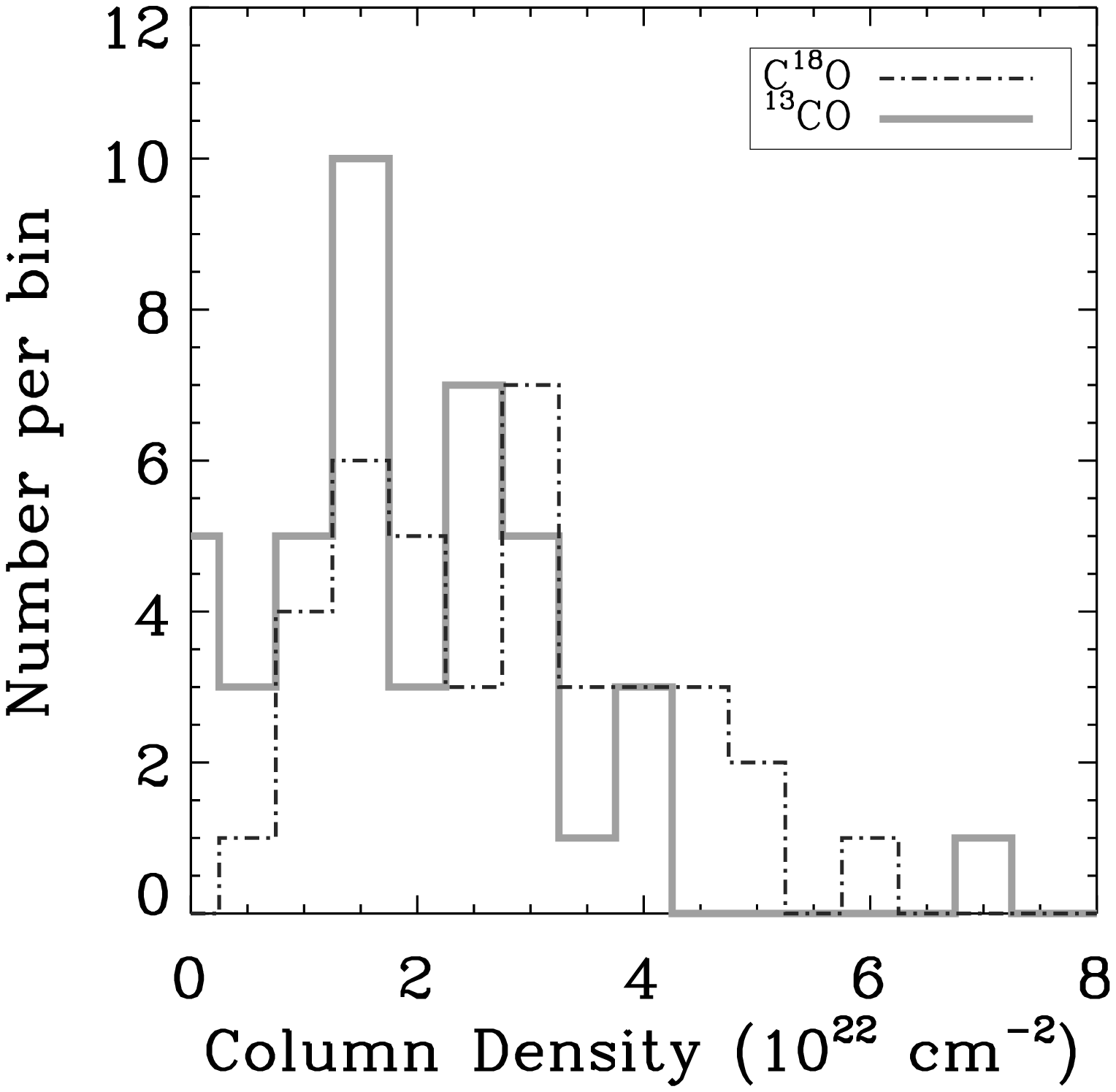}{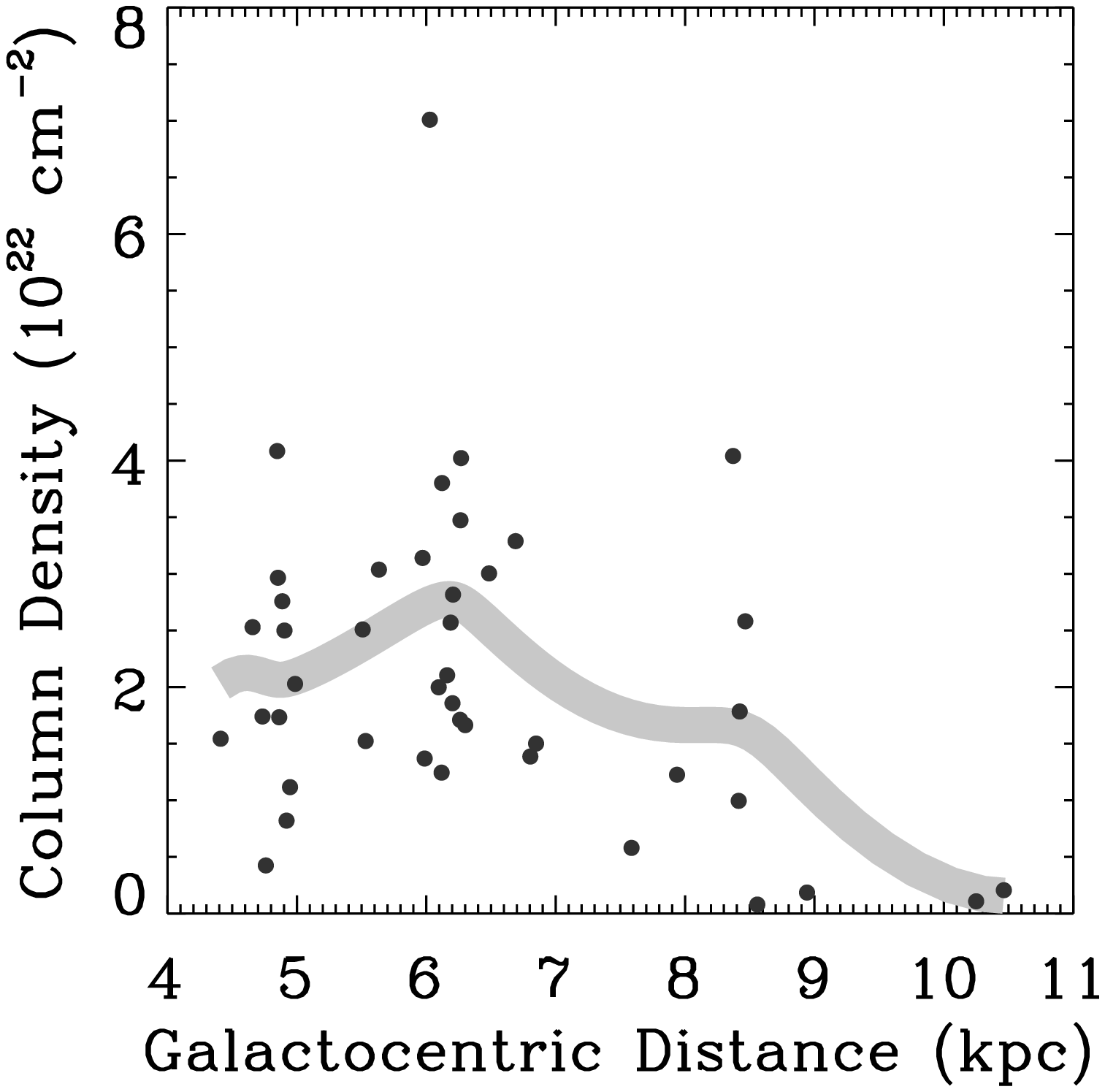}
\figcaption{
Left: Histogram of H$_2$ column densities derived from $^{13}$CO and C$^{18}$O. 
Right: Variation of column density derived from $^{13}$CO with Galactocentric distance.
\label{fig:distri_ncolden}}
\end{figure}

\begin{figure}[htbp]
\plottwo{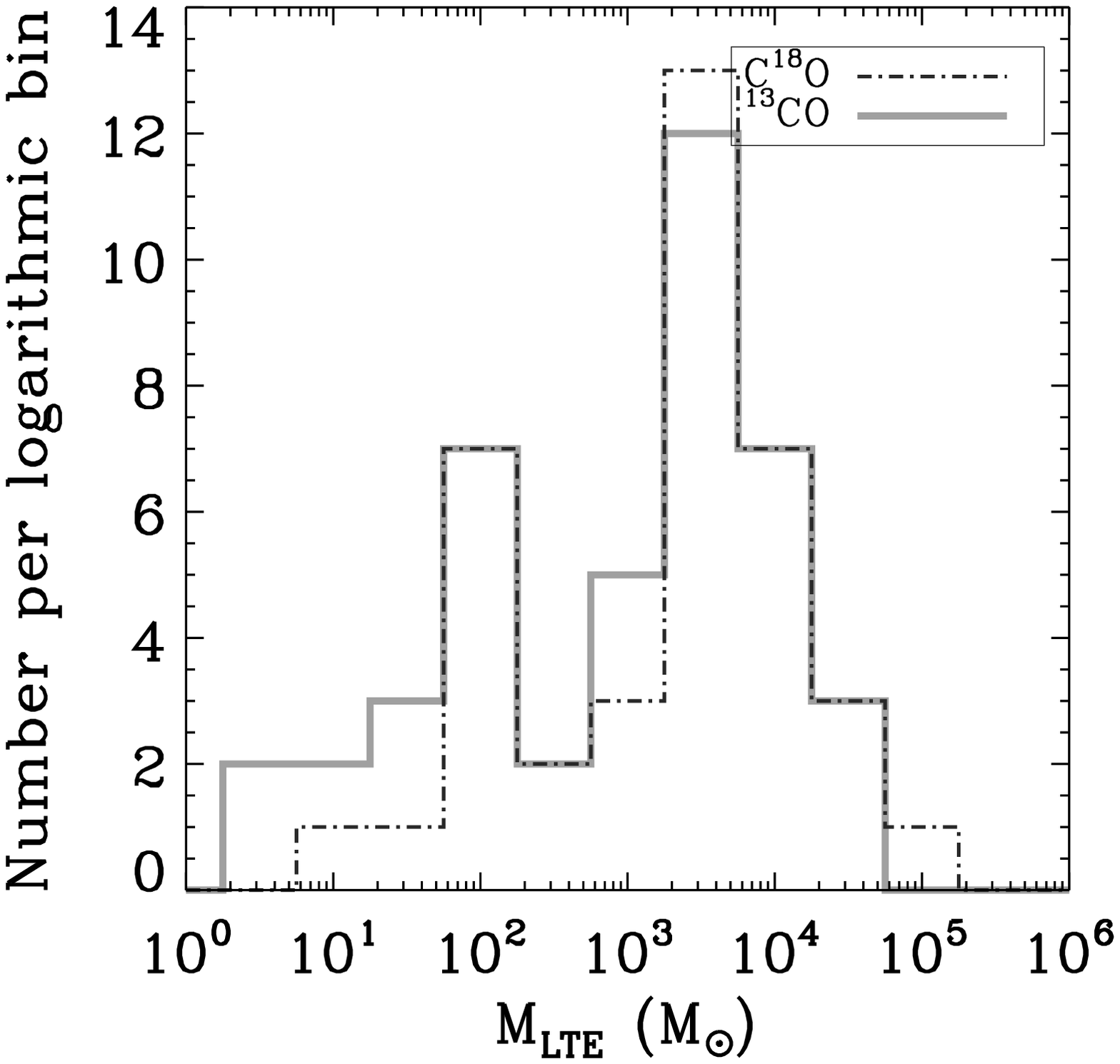}{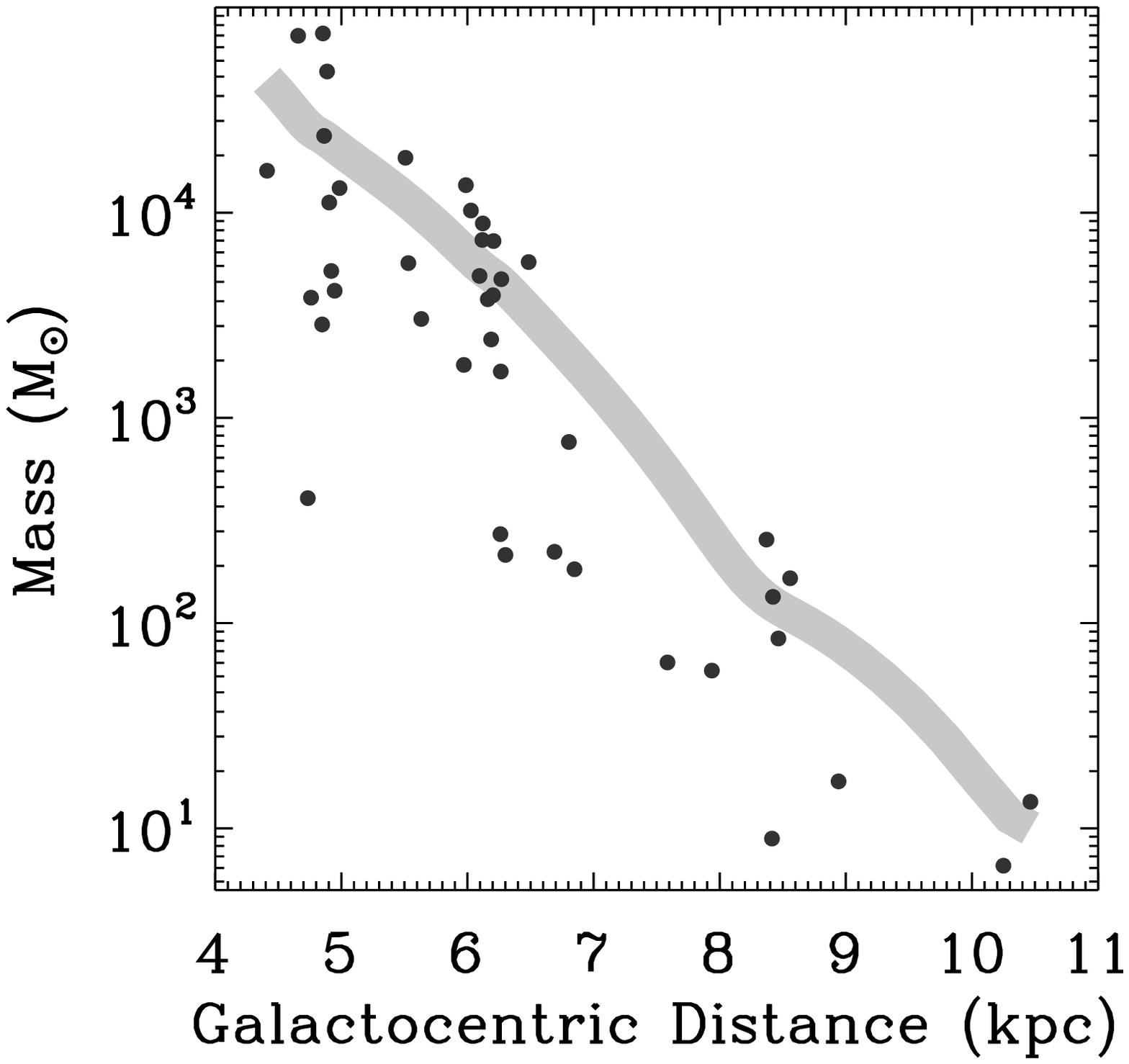}
\figcaption{
Left: Histogram of cloud LTE mass derived from $^{13}$CO and C$^{18}$O.
Right: Variation of LTE mass derived from $^{13}$CO with Galactocentric distance.
\label{fig:hmass}}
\end{figure}

We used the angular sizes in the Simon catalog as we cannot derive reasonable sizes merely from single point observations.
Linear sizes are calculated from angular sizes and kinematic distances just obtained.
Histogram of sizes, column densities, and masses are shown in Figures \ref{fig:hphysize} -- \ref{fig:hmass}.
The detection limit of column density is about $10^{21}$ cm$^{-2}$ for our survey.
The mean size of our sample is 4 pc.
The mean column density of H$_2$ derived from $^{13}$CO and C$^{18}$O are $2\times10^{22}$ cm$^{-2}$
and $3\times10^{22}$ cm$^{-2}$, while the mean LTE mass derived from $^{13}$CO and C$^{18}$O are 4500 and 7000 M$_\odot$,
and the mean volume density are 4400 and 6600 cm$^{-3}$ respectively,
which are consistent with the results of \citet{Simon2006b}.
The typical column density of our sample is not as high as that of some previous work, e.g., \citet{Carey1998a}.
This is understandable due to the limitation in the tracers we observed,
and the positions we targeted may not be the at the peaks of the clouds.
Column densities derived from $^{13}$CO tend to be underestimated due to saturation effect,
and masses derived from $^{13}$CO are usually smaller than that derived from C$^{18}$O, as the same sizes are used in calculating masses.

\begin{figure}[htbp]
\plottwo{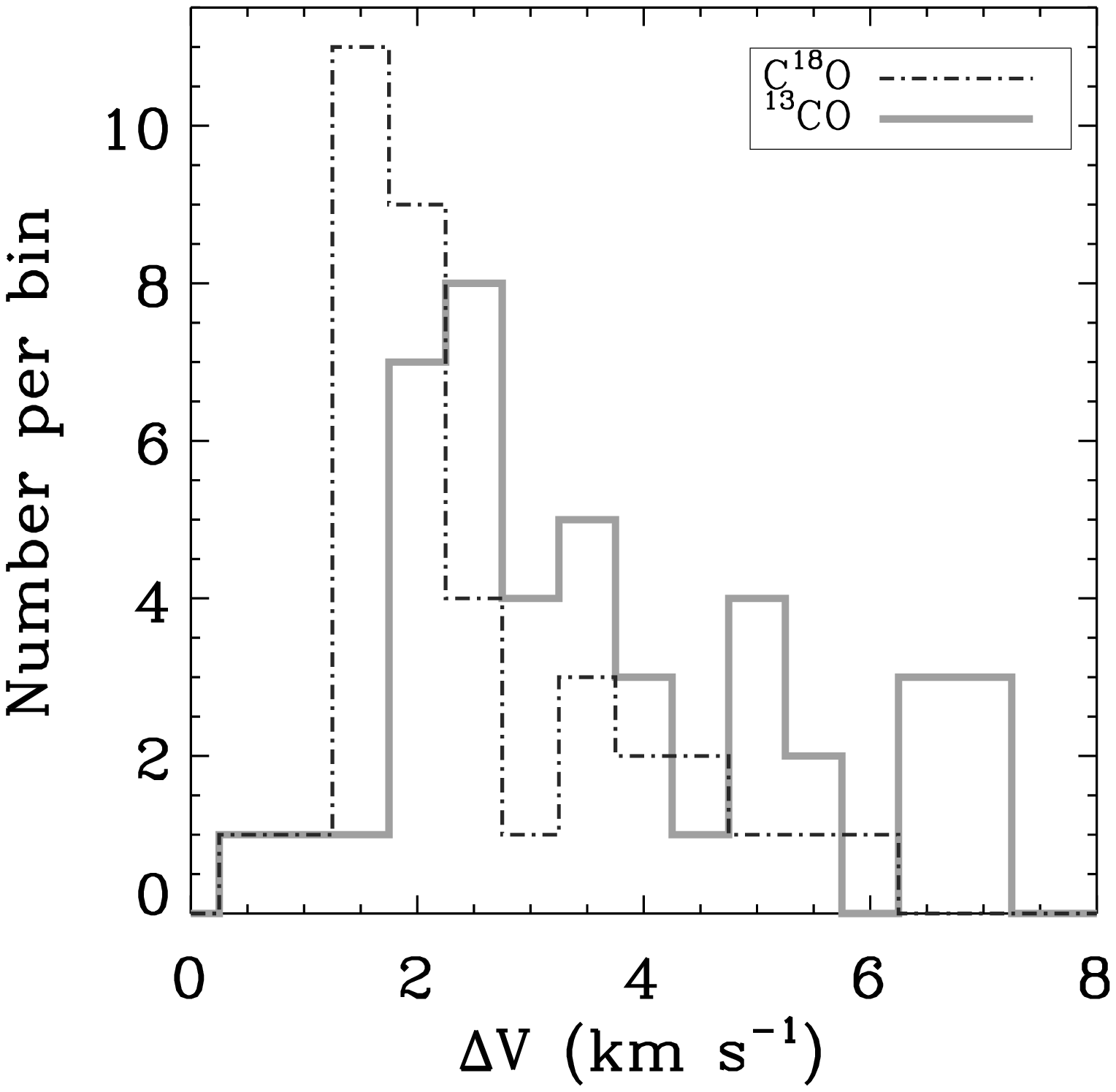}{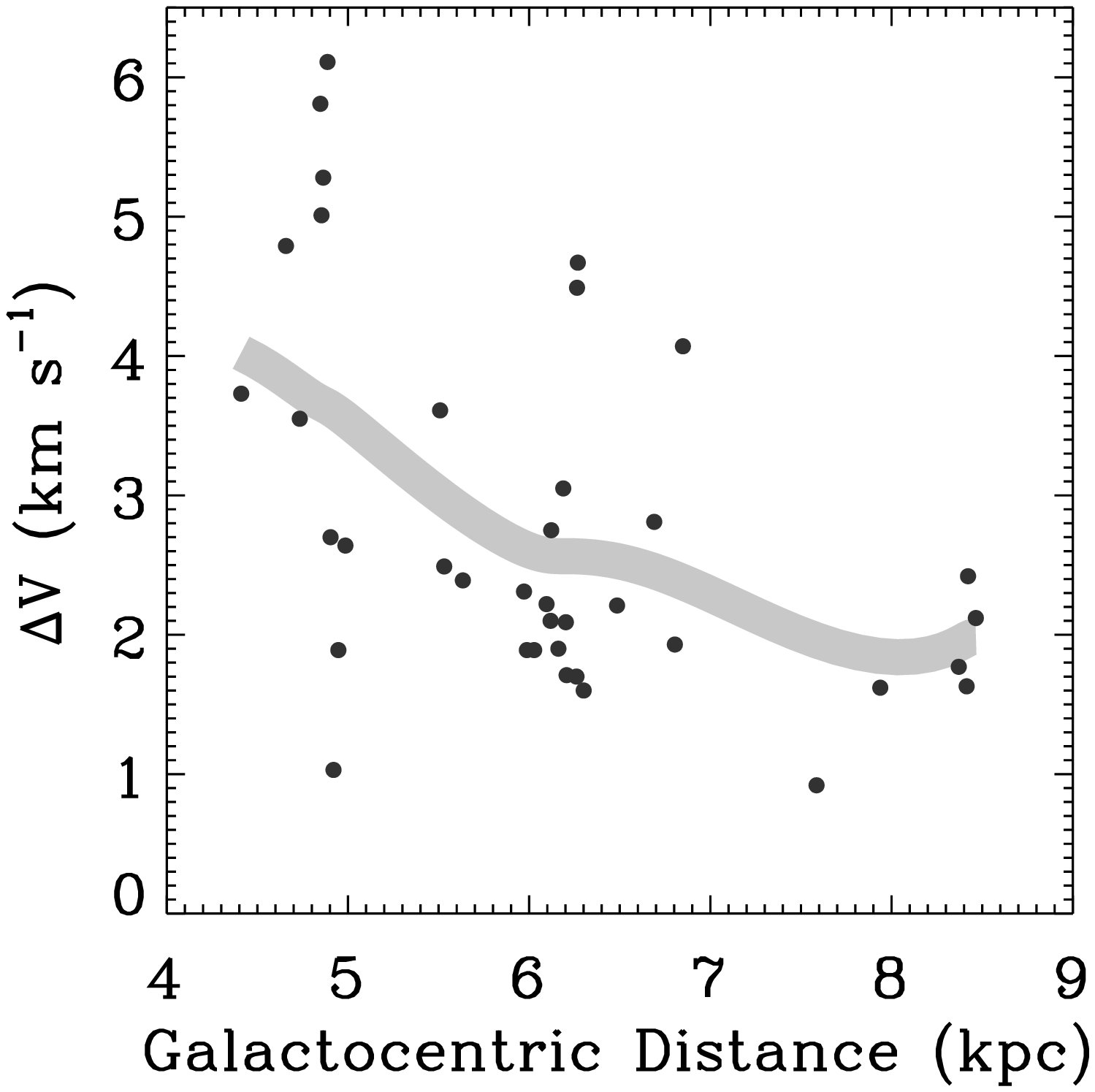}
\figcaption{Histogram of line width and the variation of C$^{18}$O line width with Galactocentric distance.
\label{fig:hdVel}}
\end{figure}

The mean line width of $^{13}$CO is 4 km s$^{-1}$ (Figure \ref{fig:hdVel}), while that of C$^{18}$O is 3 km s$^{-1}$,
much broader than that of some visually opaque regions \citep{Myers1983a}.
As finite optical depth can broaden the line profile, we attempted to make correction to the line width.
Assuming Gaussian profile for the distribution of optical depth with respect to velocity,
then the observed line width is related to the ``true'' line width by
$$\rm \frac{\Delta V_{\it line}}{\Delta V_{\it true}}=\sqrt{\frac{\ln\left[\tau_0/\ln\left(2/\left(1+e^{-\tau_0}\right)\right)\right]}{\ln2}}$$
where $\tau_0$ is the peak optical depth.
After converting the line widths into ``true'' widths,
the ``true'' widths of $^{13}$CO and C$^{18}$O become closer, although those of $^{13}$CO are still about 1.2 -- 2 times broader.
There are two possibilities.
Maybe the intrinsic line width of $^{13}$CO and C$^{18}$O is the same if their abundance ratio is constant throughout the cloud;
the reason why those of $^{13}$CO are broader might be that
the $^{13}$CO lines are rather saturated and the optical depths derived here tend to be underestimated,
which makes the previous modification insufficient.
However, it is more likely that the C$^{18}$O line widths are intrinsically narrower.
As the C$^{18}$O lines tend to trace denser regions, 
the kinematic structure of the cloud might cause this kind of discrepancy in velocity dispersion;
it is possibile that in the central region the kinematic energy is dissipated to give way to the further collapse or accretion motion.
As is speculated by \citet{Myers1983b}, turbulence in the macro-scale is primarily caused by collisions and drag,
while there is no such source in the micro-scale, and turbulence in this scale can continuously decay.
However, our current survey is insufficient to give any definite clue to this, and detailed map of central regions in dense gas tracers are needed.

\clearpage

\clearpage

\begin{figure}[htbp]
\plottwo{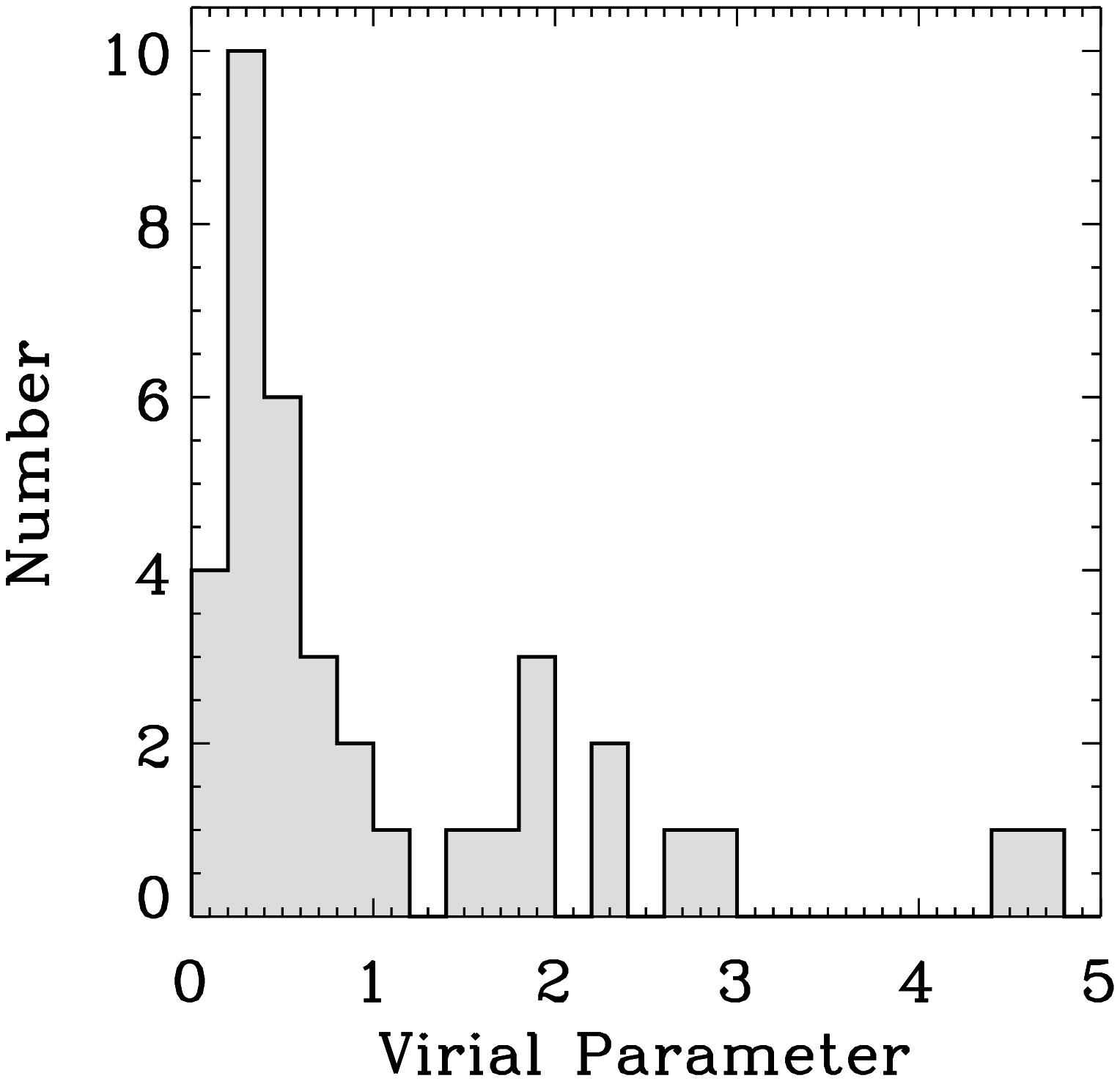}{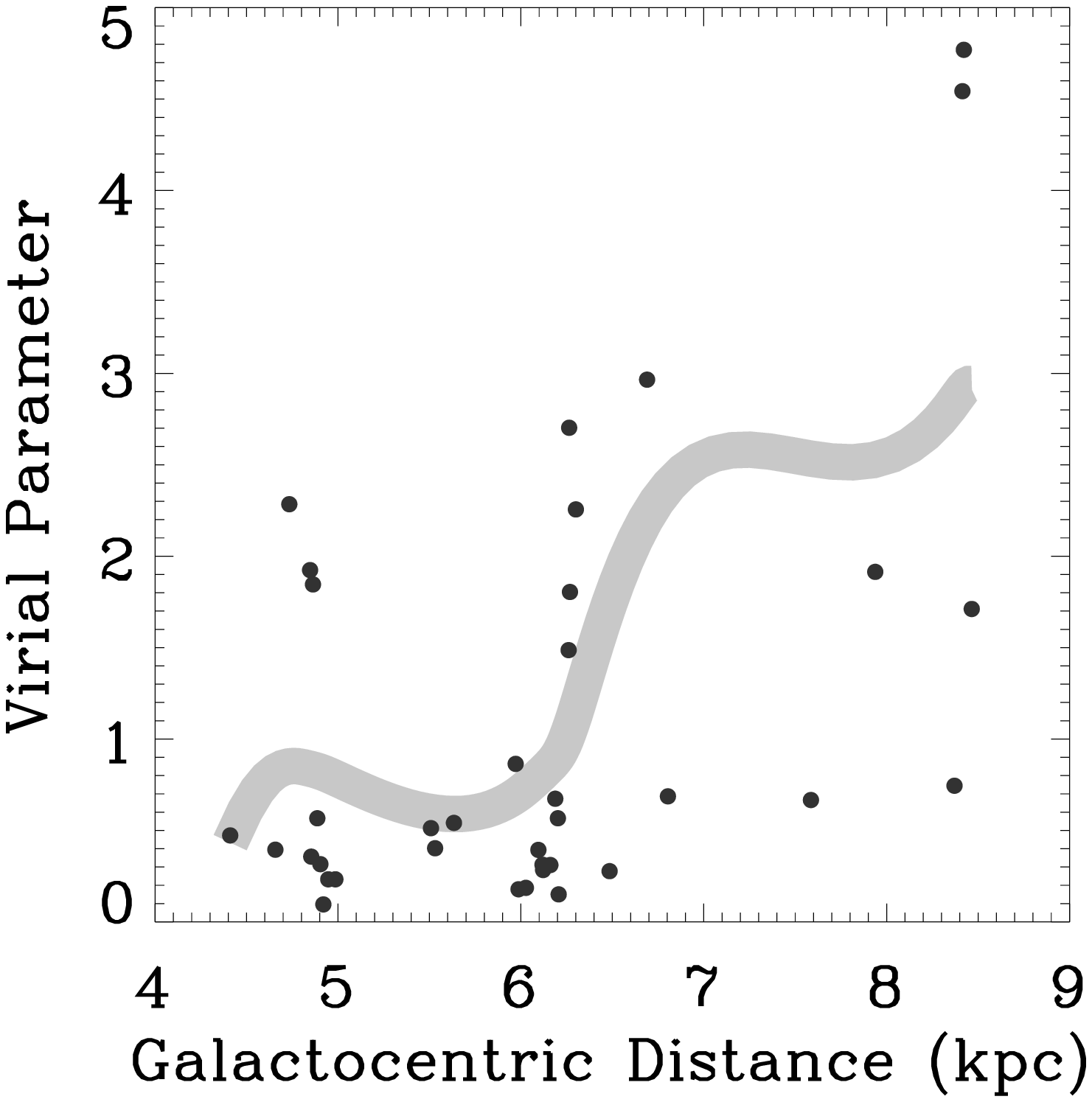}
\figcaption{Histogram of virial parameters  and its variation with Galactocentric distance.
\label{fig:hturb_grav}}
\end{figure}

\begin{figure}[htbp]
\plottwo{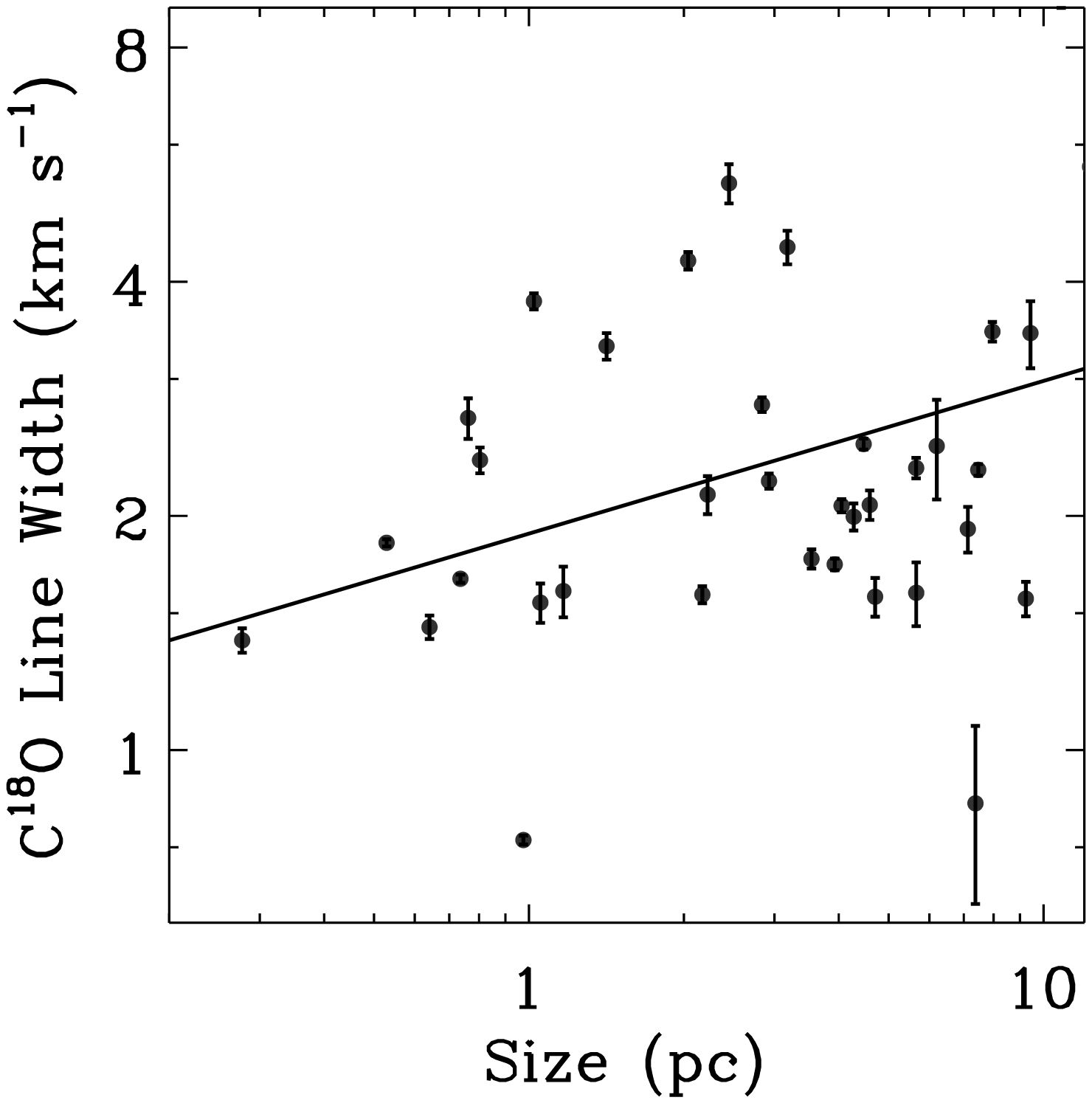}{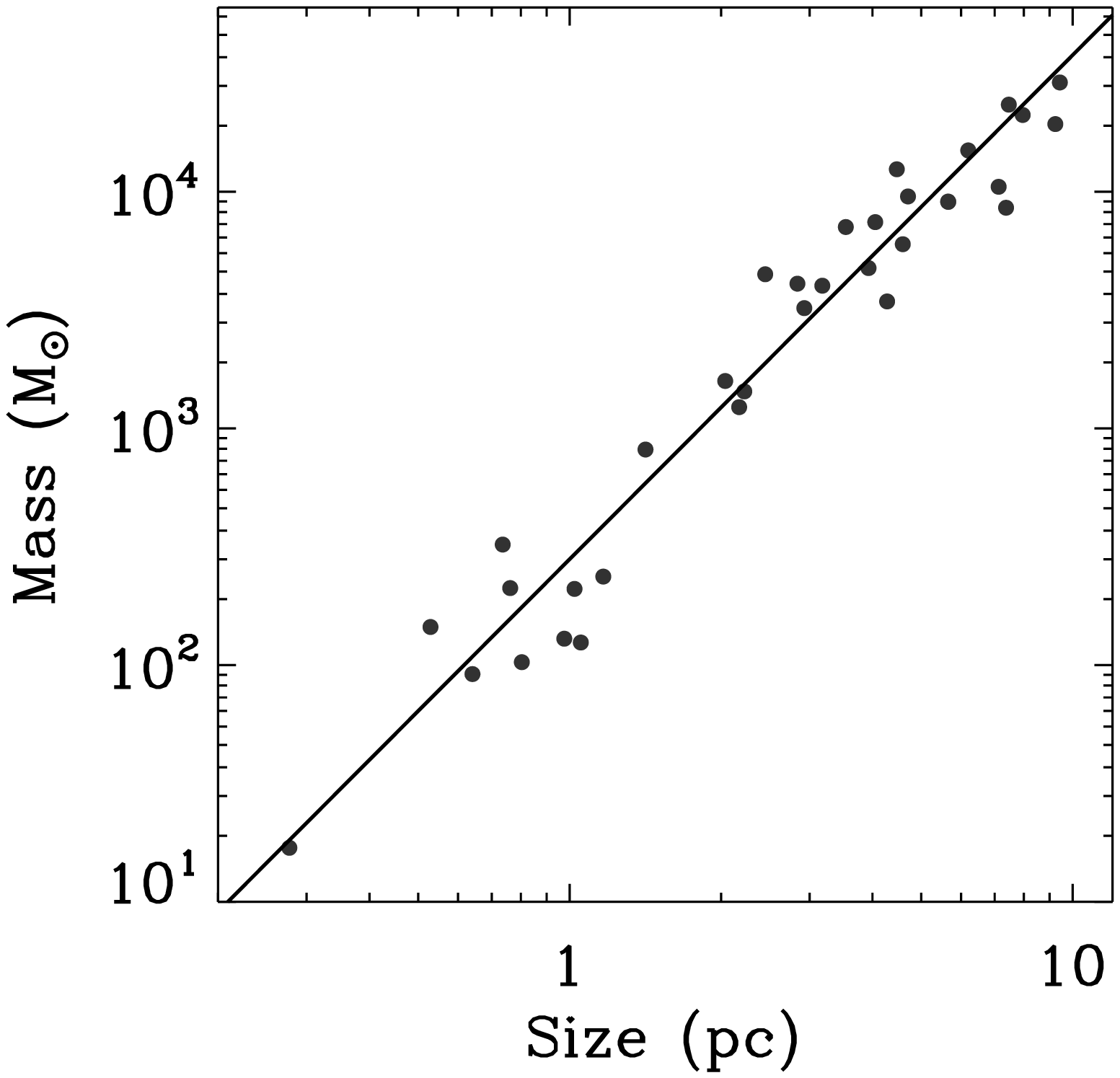}
\figcaption{Plot of sizes versus line widths and masses. Power law fittings are overplotted. \label{fig:virial_size}}
\end{figure}

The line widths mainly originate from turbulent motions,
and we may compare the kinematic energy density with the gravitational energy density.
Assuming uniform spheres, the ratio of twice the kinematic energy density to gravitational energy density can be expressed as
$$\alpha=\frac{5\sigma^2R}{G M},$$
which is denominated virial parameter by some authors \citep{Bertoldi1992a, Krumholz2005a}.
Here R is the radius of the sphere, $\sigma$ is the average total one-dimensional velocity dispersion,
which is related to the Gaussian fit FWHM by $\sigma$$=$${\rm FWHM}$$/$$\sqrt{8\ln2}$.
Histogram of $\alpha$ for those clouds detected in C$^{18}$O is shown in Figure \ref{fig:hturb_grav}.
It has an average of 1.3 and median of 0.6, which means these clouds are near virial equilibrium.
However, uncertainties in our assumption of size, geometry and matter distribution
(together with observational uncertainties and fluctuations in individual targets)
limit the validity of our result. 
As in this paper the sizes for the IRDCs are calculated using the kinematic distances we derived herein and the angular sizes in the Simon catalog,
it is probable that sizes of the molecular clouds associated with the IRDCs are underestimated, 
and the virial parameters for the clouds may actually be smaller, which means collapse motions are inevitable in the overall scale.
In Figure \ref{fig:virial_size} we plot the sizes versus the line widths and masses,
overlaid with power law fittings which can be expressed as
$$\Delta V \propto R^{0.2\pm0.07},$$
$$M \propto R^{2.1\pm0.07}.$$
These correlations imply that $\alpha\propto R^{-0.7}$, $\alpha\propto M^{-0.3}$.
The fact that the mass being approximately proportional to R$^2$ is straightforward if 
the column densities are not correlated with sizes, which is actually the case for our survey.
Being in virial equilibrium is a natural explanation for this.
Another possibility is that the usual LTE method in determining the optical depth and the column density can easily be affected by observational noises,
and this might obscure the correlation of column density with radius.
That the velocity widths only weakly correlate with the sizes and the index of the power law relation is smaller than the typical value 0.5 of
some previous studies \citep{Myers1983b, Dame1986a, Solomon1987a, Goodman1993a} is probably due to the fact that
sizes of the extinction features cannot all be very accurate as for many IRDCs the background emission is not smooth and
the Gassian fit procedure in deriving the sizes might be problematic \citep{Simon2006a}.
Besides this, the sizes of the extinction features may not necessarily correlate with the sizes of the molecular clouds.

\clearpage

\clearpage

\subsection{Correlation between the MSX data and the molecular line data}
\begin{figure}[htbp]
\plotone{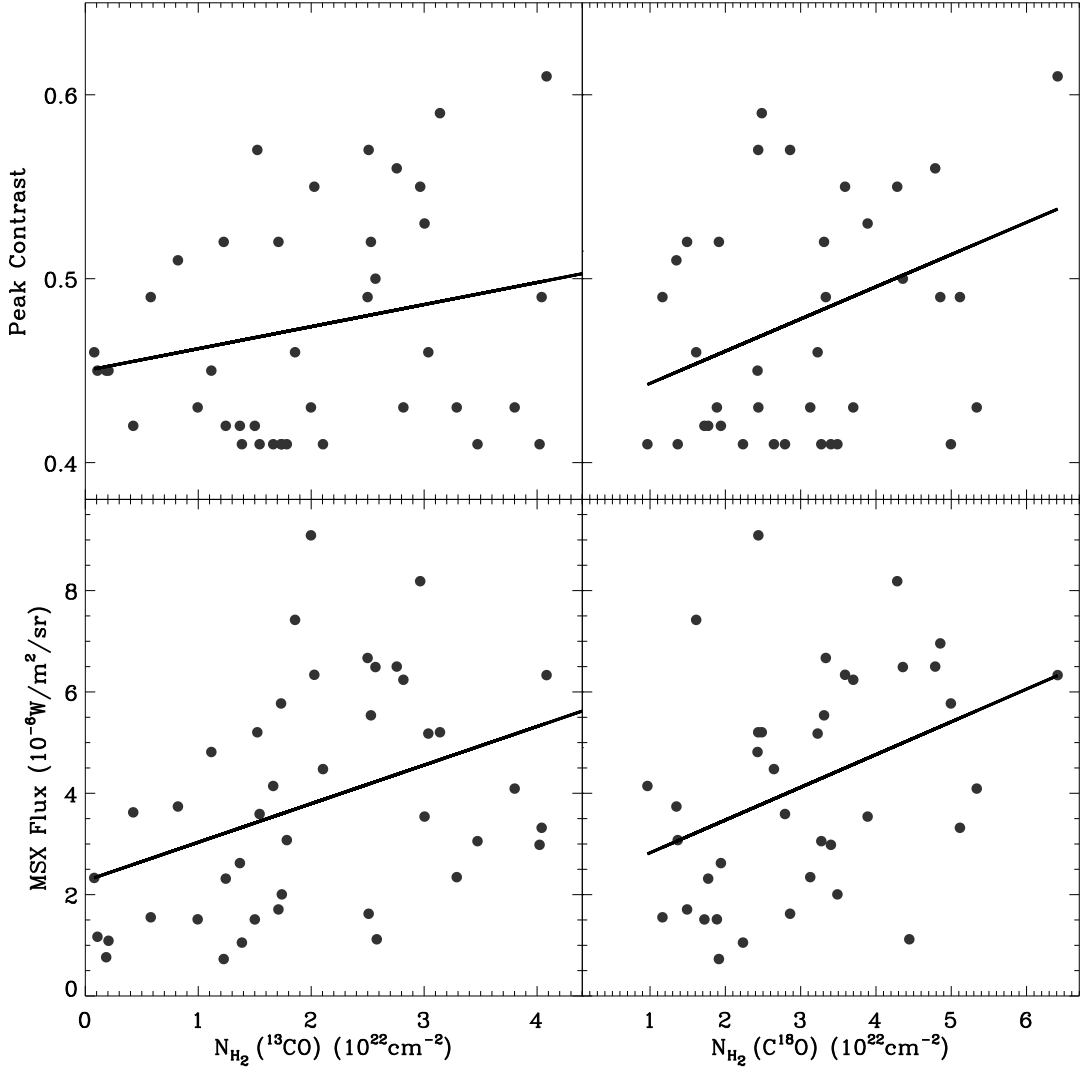}
\figcaption{Correlation between the MSX flux as well as the peak contrast and the H$_2$ column density derived from $^{13}$CO and C$^{18}$O. \label{fig:MSXflux_ncold}}
\end{figure}

As our sources are selected from the MSX data, it is natural to compare our observation with the MSX images.
Initially we expected that the column density should be positively correlated with the degree of extinction of the IRDCs,
which is properly described by the contrast parameter in the Simon catalog.
However, the correlation between the contrast and the column density turns out to be rather weak, although it is positive.
The MSX flux is also positively correlated with the column density.
These correlations can be described by the following formulae (Figure \ref{fig:MSXflux_ncold}):
$$\rm
\mbox{Contrast} = 0.45 + 0.012 \times \left(\frac{N_{H_2}({}^{13}CO)}{10^{22}\ \mbox{cm}^{-2}}\right)
$$ $$\rm
\mbox{Contrast} = 0.43 + 0.018 \times \left(\frac{N_{H_2}(C^{18}O)}{10^{22}\ \mbox{cm}^{-2}}\right)
$$ $$\rm
\left(\frac{\mbox{F(8.3 $\mu$m)}}{10^{-6}\ W/m^2/sr}\right) = 2.3 + 0.8 \times \left(\frac{N_{H_2}({}^{13}CO)}{10^{22}\ \mbox{cm}^{-2}}\right)
$$ $$\rm
\left(\frac{\mbox{F(8.3 $\mu$m)}}{10^{-6}\ W/m^2/sr}\right) = 2.2 + 0.6 \times \left(\frac{N_{H_2}(C^{18}O)}{10^{22}\ \mbox{cm}^{-2}}\right)
$$
Here the contrasts are the peak contrasts taken from the Simon catalog, 
and the MSX fluxes are averaged over a $10''\times10''$ region near the IRDC peaks.

The correlation between MSX brightness and the column density is plausible,
as brighter infrared background (or neighbor) implies more drastic activities,
which might trigger gravitational instability and lead to condensation of matter;
also bright infrared emission usually comes from inner part of the Galaxy,
where large amounts of molecular clouds reside.
As for the insignificant correlation between peak contrast and column density,
foreground contamination, weak and noisy background emission, variability of dust-to-gas ratio,
and/or inaccurate contrast in the Simon catalog resulting from complex environment of some IRDCs can be the reason.
Also the narrow contrast range of our sample might make the correlation look insignificant.
For high extinction, the contrast is not a good approximation of optical depth.
It is also possible that some IRDCs may be embedded in larger or giant molecular clouds,
and the CO lines may actually trace these environmental clouds, being insensitive to the denser IRDCs.
Furthermore, some IRDCs actually seem to be gaps in bright emission regions, rather than extinction features;
this kind of IRDCs are usually  morphologically connected to the dark vacuum outer space above or below the Galactic disk.

\clearpage
\clearpage

\section{Conclusions}
We observed 61 MSX IRDCs in three CO isotope lines with coordinates mainly taken from the Simon catalog.
For most sources, excitation temperatures, distances, and column densities are derived.
The Galactic distribution of IRDCs is consistent with the 5 kpc molecular ring picture, 
while showing some traces of spiral pattern, 
but a larger and more uniform survey is needed to determine whether this feature is true,
and maybe independent method to determine the distances other than the rotation curve method needs to be implemented.
Sizes are estimated using angular sizes from the Simon catalog, and LTE masses are calculated subsequently.
Typical size of the IRDC is several $pc$s, typical column density is several $10^{22}$ cm$^{-2}$, 
typical density is about 5000 cm$^{-3}$, and typical mass is about 5000 M$_\odot$, 
which are similar to that of star forming clumps.
Many sources are significantly saturated in $^{13}$CO.
The abundance ratio of $^{13}$CO to C$^{18}$O in IRDCs is similar to that of typical molecular clouds.
The column density of IRDCs only weakly correlate with their peak contrast,
which might indicate that the peak contrast of some IRDCs are inaccurate due to noisy background,
as well as bright foreground and complex neighbourhood.
As giant molecular clouds have typical size of tens of pc, typical mass of $10^5$ -- $10^6$ M$_\odot$, 
and typical density of hundreds of cm$^{-3}$ \citep{Solomon1979, Sanders1985a}, 
while Bok globules have typical size of several 0.1 pc, typical mass of 2 -- 100 M$_\odot$, 
and typical density of $10^4$ -- $10^5$ cm$^{-3}$ \citep{Launhardt1997a}, 
IRDC seems to be an intermediate class between these two species.

Many questions can be raised.
Are all IRDCs in the same stage of evolution?
Or more basically, are all IRDCs intrinsically the same? Do they fall into several different species?
What's the relationship between the IRDCs and their neighbor bright regions?
Are they compressed by their neighbors?
What role do IRDCs play in the formation of massive stars?
Whether there are pre-stellar objects in the central parts of IRDCs already?
Furthermore, it is not clear whether the sequence of GMC, IRDC, and Bok globule is an evolution sequence,
and, if so, what's the underlying mechanism?
Detailed mapping of some of these sources in a variety of tracers is necessary to clarify these issues.

\acknowledgments
This work made use of data products from the Midcourse Space Experiment.
We are grateful to the staff members of the Delingha radio telescope for their help during the observations.
We thank the anonymous referee for his/her valuable comments and suggestions to the draft of this paper.
We also thank R. Simon for valuable clarification of some issues regarding their paper.
This work was supported by NSFC through grant 10621303 and Ministry of Science and Technology of China through grant 2007CB815406.

\end{document}